\documentclass[longauth]{aa} %

\usepackage{xcolor}
\newcommand{\todo}[1]{}
\renewcommand{\todo}[1]{{\color{red} TODO: {#1}}}

\usepackage{euclid}

\usepackage{natbib}
\bibpunct{(}{)}{;}{a}{}{,} %

\usepackage{scalerel}

\usepackage[pdfencoding=auto,psdextra]{hyperref}
\hypersetup{
    colorlinks=true,
    linkcolor=blue,
    filecolor=magenta,      
    urlcolor=blue,
    citecolor=blue
}
\makeatletter
\renewcommand*\aa@pageof{, page \thepage{} of \pageref*{LastPage}}
\makeatother

\usepackage[utf8]{inputenc}

\usepackage[switch, modulo]{lineno}

\usepackage{graphicx}
\usepackage{txfonts}
\usepackage{textcomp} %
\usepackage{comment}

\usepackage{csquotes}

\graphicspath{ {./Images/} }

\begin{document} 

\title{\Euclid: Identification of asteroid streaks in simulated images using deep learning\thanks{This paper is published on behalf of the Euclid Consortium.}}

\newcommand{\orcid}[1]{} %
\author{M.~P{\"o}ntinen\orcid{0000-0001-5442-2530}$^{1}$\thanks{\email{mikko.pontinen@helsinki.fi}}, M.~Granvik\orcid{0000-0002-5624-1888}$^{1,2}$, A.~A.~Nucita$^{3,4,5}$, L.~Conversi\orcid{0000-0002-6710-8476}$^{6,7}$, B.~Altieri\orcid{0000-0003-3936-0284}$^{7}$, B.~Carry$^{8}$, C.~M.~O'Riordan$^{9}$, D.~Scott\orcid{0000-0002-6878-9840}$^{10}$, N.~Aghanim$^{11}$, A.~Amara$^{12}$
% ...
}

\institute{$^{1}$ Department of Physics, P.O. Box 64, 00014 University of Helsinki, Finland\\
$^{2}$ Division of Space Technology, Lule\aa{} University of Technology, Box 848, 98128 Kiruna, Sweden\\
$^{3}$ Department of Mathematics and Physics E. De Giorgi, University of Salento, Via per Arnesano, CP-I93, 73100, Lecce, Italy\\
$^{4}$ INAF-Sezione di Lecce, c/o Dipartimento Matematica e Fisica, Via per Arnesano, 73100, Lecce, Italy\\
$^{5}$ INFN, Sezione di Lecce, Via per Arnesano, CP-193, 73100, Lecce, Italy\\
$^{6}$ European Space Agency/ESRIN, Largo Galileo Galilei 1, 00044 Frascati, Roma, Italy\\
$^{7}$ ESAC/ESA, Camino Bajo del Castillo, s/n., Urb. Villafranca del Castillo, 28692 Villanueva de la Ca\~nada, Madrid, Spain\\
$^{8}$ Universit\'e C\^{o}te d'Azur, Observatoire de la C\^{o}te d'Azur, CNRS, Laboratoire Lagrange, Bd de l'Observatoire, CS 34229, 06304 Nice cedex 4, France\\
$^{9}$ Max-Planck-Institut f\"ur Astrophysik, Karl-Schwarzschild Str. 1, 85741 Garching, Germany\\
$^{10}$ Departement of Physics and Astronomy, University of British Columbia, Vancouver, BC V6T 1Z1, Canada\\
$^{11}$ Universit\'e Paris-Saclay, CNRS, Institut d'astrophysique spatiale, 91405, Orsay, France\\
$^{12}$ Institute of Cosmology and Gravitation, University of Portsmouth, Portsmouth PO1 3FX, UK\\
$^{13}$ Institut f\"ur Theoretische Physik, University of Heidelberg, Philosophenweg 16, 69120 Heidelberg, Germany\\
$^{14}$ INAF-Osservatorio di Astrofisica e Scienza dello Spazio di Bologna, Via Piero Gobetti 93/3, 40129 Bologna, Italy\\
$^{15}$ Dipartimento di Fisica e Astronomia "Augusto Righi" - Alma Mater Studiorum Universit\`a di Bologna, via Piero Gobetti 93/2, 40129 Bologna, Italy\\
$^{16}$ INFN-Sezione di Bologna, Viale Berti Pichat 6/2, 40127 Bologna, Italy\\
$^{17}$ INAF-Osservatorio Astrofisico di Torino, Via Osservatorio 20, 10025 Pino Torinese (TO), Italy\\
$^{18}$ Dipartimento di Fisica, Universit\`a di Genova, Via Dodecaneso 33, 16146, Genova, Italy\\
$^{19}$ INFN-Sezione di Genova, Via Dodecaneso 33, 16146, Genova, Italy\\
$^{20}$ Department of Physics "E. Pancini", University Federico II, Via Cinthia 6, 80126, Napoli, Italy\\
$^{21}$ INAF-Osservatorio Astronomico di Capodimonte, Via Moiariello 16, 80131 Napoli, Italy\\
$^{22}$ Dipartimento di Fisica, Universit\`a degli Studi di Torino, Via P. Giuria 1, 10125 Torino, Italy\\
$^{23}$ INFN-Sezione di Torino, Via P. Giuria 1, 10125 Torino, Italy\\
$^{24}$ INAF-IASF Milano, Via Alfonso Corti 12, 20133 Milano, Italy\\
$^{25}$ Institut de F\'{i}sica d'Altes Energies (IFAE), The Barcelona Institute of Science and Technology, Campus UAB, 08193 Bellaterra (Barcelona), Spain\\
$^{26}$ Port d'Informaci\'{o} Cient\'{i}fica, Campus UAB, C. Albareda s/n, 08193 Bellaterra (Barcelona), Spain\\
$^{27}$ INAF-Osservatorio Astronomico di Roma, Via Frascati 33, 00078 Monteporzio Catone, Italy\\
$^{28}$ INFN section of Naples, Via Cinthia 6, 80126, Napoli, Italy\\
$^{29}$ Dipartimento di Fisica e Astronomia "Augusto Righi" - Alma Mater Studiorum Universit\`a di Bologna, Viale Berti Pichat 6/2, 40127 Bologna, Italy\\
$^{30}$ Centre National d'Etudes Spatiales -- Centre spatial de Toulouse, 18 avenue Edouard Belin, 31401 Toulouse Cedex 9, France\\
$^{31}$ Institut national de physique nucl\'eaire et de physique des particules, 3 rue Michel-Ange, 75794 Paris C\'edex 16, France\\
$^{32}$ Institute for Astronomy, University of Edinburgh, Royal Observatory, Blackford Hill, Edinburgh EH9 3HJ, UK\\
$^{33}$ University of Lyon, Univ Claude Bernard Lyon 1, CNRS/IN2P3, IP2I Lyon, UMR 5822, 69622 Villeurbanne, France\\
$^{34}$ Institute of Physics, Laboratory of Astrophysics, Ecole Polytechnique F\'ed\'erale de Lausanne (EPFL), Observatoire de Sauverny, 1290 Versoix, Switzerland\\
$^{35}$ Mullard Space Science Laboratory, University College London, Holmbury St Mary, Dorking, Surrey RH5 6NT, UK\\
$^{36}$ Departamento de F\'isica, Faculdade de Ci\^encias, Universidade de Lisboa, Edif\'icio C8, Campo Grande, PT1749-016 Lisboa, Portugal\\
$^{37}$ Instituto de Astrof\'isica e Ci\^encias do Espa\c{c}o, Faculdade de Ci\^encias, Universidade de Lisboa, Campo Grande, 1749-016 Lisboa, Portugal\\
$^{38}$ Department of Astronomy, University of Geneva, ch. d'Ecogia 16, 1290 Versoix, Switzerland\\
$^{39}$ INFN-Padova, Via Marzolo 8, 35131 Padova, Italy\\
$^{40}$ Universit\'e Paris-Saclay, Universit\'e Paris Cit\'e, CEA, CNRS, Astrophysique, Instrumentation et Mod\'elisation Paris-Saclay, 91191 Gif-sur-Yvette, France\\
$^{41}$ INAF-Osservatorio Astronomico di Trieste, Via G. B. Tiepolo 11, 34143 Trieste, Italy\\
$^{42}$ Aix-Marseille Universit\'e, CNRS/IN2P3, CPPM, Marseille, France\\
$^{43}$ INAF-Osservatorio Astronomico di Padova, Via dell'Osservatorio 5, 35122 Padova, Italy\\
$^{44}$ Institute of Theoretical Astrophysics, University of Oslo, P.O. Box 1029 Blindern, 0315 Oslo, Norway\\
$^{45}$ Jet Propulsion Laboratory, California Institute of Technology, 4800 Oak Grove Drive, Pasadena, CA, 91109, USA\\
$^{46}$ von Hoerner \& Sulger GmbH, Schlo{\ss}Platz 8, 68723 Schwetzingen, Germany\\
$^{47}$ Technical University of Denmark, Elektrovej 327, 2800 Kgs. Lyngby, Denmark\\
$^{48}$ Cosmic Dawn Center (DAWN), Denmark\\
$^{49}$ Max-Planck-Institut f\"ur Astronomie, K\"onigstuhl 17, 69117 Heidelberg, Germany\\
$^{50}$ Universit\"ats-Sternwarte M\"unchen, Fakult\"at f\"ur Physik, Ludwig-Maximilians-Universit\"at M\"unchen, Scheinerstrasse 1, 81679 M\"unchen, Germany\\
$^{51}$ Universit\'e de Gen\`eve, D\'epartement de Physique Th\'eorique and Centre for Astroparticle Physics, 24 quai Ernest-Ansermet, CH-1211 Gen\`eve 4, Switzerland\\
$^{52}$ Helsinki Institute of Physics, Gustaf H{\"a}llstr{\"o}min katu 2, University of Helsinki, Helsinki, Finland\\
$^{53}$ NOVA optical infrared instrumentation group at ASTRON, Oude Hoogeveensedijk 4, 7991PD, Dwingeloo, The Netherlands\\
$^{54}$ Universit\"at Bonn, Argelander-Institut f\"ur Astronomie, Auf dem H\"ugel 71, 53121 Bonn, Germany\\
$^{55}$ Department of Physics, Institute for Computational Cosmology, Durham University, South Road, DH1 3LE, UK\\
$^{56}$ Universit\'e Paris Cit\'e, CNRS, Astroparticule et Cosmologie, 75013 Paris, France\\
$^{57}$ University of Applied Sciences and Arts of Northwestern Switzerland, School of Engineering, 5210 Windisch, Switzerland\\
$^{58}$ Institut d'Astrophysique de Paris, 98bis Boulevard Arago, 75014, Paris, France\\
$^{59}$ Institut d'Astrophysique de Paris, UMR 7095, CNRS, and Sorbonne Universit\'e, 98 bis boulevard Arago, 75014 Paris, France\\
$^{60}$ CEA Saclay, DFR/IRFU, Service d'Astrophysique, Bat. 709, 91191 Gif-sur-Yvette, France\\
$^{61}$ European Space Agency/ESTEC, Keplerlaan 1, 2201 AZ Noordwijk, The Netherlands\\
$^{62}$ Kapteyn Astronomical Institute, University of Groningen, PO Box 800, 9700 AV Groningen, The Netherlands\\
$^{63}$ Leiden Observatory, Leiden University, Niels Bohrweg 2, 2333 CA Leiden, The Netherlands\\
$^{64}$ Department of Physics and Astronomy, University of Aarhus, Ny Munkegade 120, DK-8000 Aarhus C, Denmark\\
$^{65}$ Universit\'e Paris-Saclay, Universit\'e Paris Cit\'e, CEA, CNRS, AIM, 91191, Gif-sur-Yvette, France\\
$^{66}$ Space Science Data Center, Italian Space Agency, via del Politecnico snc, 00133 Roma, Italy\\
$^{67}$ Max Planck Institute for Extraterrestrial Physics, Giessenbachstr. 1, 85748 Garching, Germany\\
$^{68}$ Dipartimento di Fisica e Astronomia "G. Galilei", Universit\`a di Padova, Via Marzolo 8, 35131 Padova, Italy\\
$^{69}$ Dipartimento di Fisica e Astronomia, Universit\`a di Bologna, Via Gobetti 93/2, 40129 Bologna, Italy\\
$^{70}$ Departamento de F\'isica, FCFM, Universidad de Chile, Blanco Encalada 2008, Santiago, Chile\\
$^{71}$ Institut d'Estudis Espacials de Catalunya (IEEC), Carrer Gran Capit\'a 2-4, 08034 Barcelona, Spain\\
$^{72}$ Institut de Ciencies de l'Espai (IEEC-CSIC), Campus UAB, Carrer de Can Magrans, s/n Cerdanyola del Vall\'es, 08193 Barcelona, Spain\\
$^{73}$ Centro de Investigaciones Energ\'eticas, Medioambientales y Tecnol\'ogicas (CIEMAT), Avenida Complutense 40, 28040 Madrid, Spain\\
$^{74}$ Instituto de Astrof\'isica e Ci\^encias do Espa\c{c}o, Faculdade de Ci\^encias, Universidade de Lisboa, Tapada da Ajuda, 1349-018 Lisboa, Portugal\\
$^{75}$ Universidad Polit\'ecnica de Cartagena, Departamento de Electr\'onica y Tecnolog\'ia de Computadoras,  Plaza del Hospital 1, 30202 Cartagena, Spain\\
$^{76}$ Institut de Recherche en Astrophysique et Plan\'etologie (IRAP), Universit\'e de Toulouse, CNRS, UPS, CNES, 14 Av. Edouard Belin, 31400 Toulouse, France\\
$^{77}$ INFN-Bologna, Via Irnerio 46, 40126 Bologna, Italy\\
$^{78}$ Infrared Processing and Analysis Center, California Institute of Technology, Pasadena, CA 91125, USA\\
$^{79}$ Junia, EPA department, 41 Bd Vauban, 59800 Lille, France}

\date{Received 24 July 2023; accepted 25 September 2023}

  \abstract
   {The material composition of asteroids is an essential piece of knowledge in the quest to understand the formation and evolution of the Solar System. Visual to near-infrared spectra or multiband photometry is required to constrain the material composition of asteroids, but we currently have such data, especially in the near-infrared wavelengths, for only a limited number of asteroids. This is a significant limitation considering the complex orbital structures of the asteroid populations. Up to 150\,000 asteroids will be visible in the images of the upcoming ESA \textit{Euclid} space telescope, and the instruments of \textit{Euclid} will offer multiband visual to near-infrared photometry and slitless near-infrared spectra of these objects. Most of the asteroids will appear as streaks in the images. Due to the large number of images and asteroids, automated detection methods are needed. A non-machine-learning approach based on the \texttt{StreakDet} software was previously tested, but the results were not optimal for short and/or faint streaks. We set out to improve the capability to detect asteroid streaks in \textit{Euclid} images by using deep learning.
   
   We built, trained, and tested a three-step machine-learning pipeline with simulated \textit{Euclid} images. First, a convolutional neural network (CNN) detected streaks and their coordinates in full images, aiming to maximize the completeness (recall) of detections. Then, a recurrent neural network (RNN) merged snippets of long streaks detected in several parts by the CNN. Lastly, gradient-boosted trees (\texttt{XGBoost}) linked detected streaks between different \textit{Euclid} exposures to reduce the number of false positives and improve the purity (precision) of the sample.
   
    The deep-learning pipeline surpasses the completeness and reaches a similar level of purity of a non-machine-learning pipeline based on the \texttt{StreakDet} software. Additionally, the deep-learning pipeline can detect asteroids 0.25--0.5 magnitudes fainter than \texttt{StreakDet}. The deep-learning pipeline could result in a 50\% increase in the number of detected asteroids compared to the \texttt{StreakDet} software. There is still scope for further refinement, particularly in improving the accuracy of streak coordinates and enhancing the completeness of the final stage of the pipeline, which involves linking detections across multiple exposures.}


    \authorrunning{M. Pöntinen et al.}

   \maketitle

\section{Introduction}

The European Space Agency's (ESA) \textit{Euclid} space telescope is built to study the nature of dark energy, dark matter, and gravity by observing weak gravitational lensing, baryon acoustic oscillations, and redshift-space distortion \citep{amendola2018}. \textit{Euclid} belongs to the Cosmic Vision program of ESA, and it was launched on July 1, 2023. The mission operates in the second Lagrange point of the Sun-Earth system, whence it surveys approximately one-third (15\,000\,deg\textsuperscript{2}) of the sky \citep{laureijs2011}.

The telescope of \textit{Euclid} has a focal length of 24.5 meters and an aperture of 1.2 meters \citep{venancio2014}. The measuring devices of \textit{Euclid} are the visible instrument VIS and the Near-Infrared Spectrometer and Photometer (NISP), both of which have a 0.53\,deg\textsuperscript{2} field of view. The pixel scales of VIS and NISP are 0.1 and 0.3 arcseconds, respectively. VIS is a 600-megapixel visible imager with an operational wavelength range from 550 to 900\,nm \citep{cropper2018}. NISP is a double instrument, consisting of a near-infrared three-filter photometer (NISP-P) and a slitless spectrograph \citep[NISP-S;][]{maciaszek2022}. The VIS instrument can detect 0.43\,arcsec extended sources in the \IE{} band to around a magnitude of $m\textsubscript{AB} = 24.9$ \citep{cropper2018}. For NISP-P with \YE{}, \JE{}, and \HE{} filters, the detection limit is around $m\textsubscript{AB} = 24.4$ \citep{schirmer2022}, whereas the NISP-S slitless spectra have a continuum sensitivity of approximately $m\textsubscript{AB} = 21$.

\textit{Euclid} will operate in a step-and-stare mode. First, VIS and NISP-S observe an area of the sky for 565 seconds, after which NISP-P carries out three 112\,s exposures with the \JE{}, \HE{}, and \YE{} filters \citep{scaramella2022}. This exposure scheme is repeated four times for each field, and the telescope pointing direction is changed, that is, dithered slightly in between. In addition to the four main VIS images, shorter VIS exposures are taken simultaneously with the NISP observations. All in all, the observing time for each field adds up to approximately 70 minutes.

As a by-product of its cosmological measurements, \textit{Euclid} observes and measures multiband photometry of up to 150\,000 Solar System objects (SSOs) \citep{carry2018}. Most of the SSOs are asteroids, but \textit{Euclid} also observes around 5000 Kuiper belt objects (KBOs), approximately 40 comets, and potentially a few InterStellar Objects (ISOs). Due to the nature of \textit{Euclid} observations, the SSOs moving faster than around $\,5\,{\rm arcsec\,h^{-1}}$ (ranging from near-Earth asteroids to Jupiter Trojans) appear as streaks in VIS images. Approximately 90 percent of the SSOs observed by the mission belong to this group \citep{carry2018}. Because \textit{Euclid} focuses on cosmological observations, it mostly avoids Galactic latitudes lower than 30\degree{} and ecliptic latitudes lower than 15\degree{}. Therefore, the asteroids detected by \textit{Euclid} are mostly on high-inclination orbits.

Many aspects of asteroid science benefit from understanding asteroid compositions \citep{gaffey2002}. In turn, compositional modeling depends on observing the spectral energy distribution of asteroids \citep{reddy2015}. The \textit{Euclid} mission is beneficial in this regard, as it substantially increases the number of asteroids with measured multiband photometry, particularly in near-infrared wavelengths. The near-infrared photometry is particularly useful for the taxonomical classification of asteroids, thus helping in the aforementioned aspects of asteroid science \citep{demeo2009, popescu2018, mahlke2022}. Additionally, the \textit{Euclid} measurements can be used to analyze asteroid rotation periods, spin-axis orientations, and asteroid shapes, as well as to detect binary asteroids \citep{carry2018}.  The utility of orbit determination with \textit{Euclid} observations alone is limited due to the relatively short observation time per asteroid, but rudimentary orbits and inclination distributions are possible to estimate. Cross-correlating the detections with other surveys, for example, with the Vera C. Rubin Observatory, makes more accurate orbit calculations possible \citep[cf.][]{snodgrass2018}. In addition to offering valuable scientific data on SSOs, identifying the objects in \textit{Euclid} images is useful to prevent them from interfering with the cosmological data-analysis pipelines.

Most of the SSOs in \textit{Euclid} data will be previously unknown objects \citep{carry2018}, so they have to be found in the images before they can be analyzed further. A notable caveat is the start of the science operations of the Vera C. Rubin Observatory Legacy Survey of Space and Time (LSST), currently planned for early 2025. LSST will discover SSOs visible from the Southern Hemisphere to a limiting magnitude similar to \textit{Euclid}.

\citet{pontinen2020} tested the streak-detection software \texttt{StreakDet}, developed by \citet{virtanen2016}, to detect SSOs in simulated \textit{Euclid} images. Overall, the results were good, but there was room for improvement in detecting short (shorter than approximately 13 pixels, corresponding to 9$\,{\rm arcsec\,h^{-1}}$) and/or faint (fainter than magnitude 23--24, depending on streak length) streaks. Nucita et al. (in prep.) developed a complementary pipeline for short streaks. On the other hand, \citet{lieu2019} studied the use of convolutional neural networks (CNNs) to find asteroids in \textit{Euclid} images. \citet{duev2019b}, \citet{duev2019}, and \citet{wang2022} have developed CNNs to reduce the number of false-positive asteroid streaks in Zwicky Transient Facility data, \citet{parfeni2020} have tested CNNs for asteroid detection with \textit{Hubble} images, and \citet{rabeendran2021} have tested CNNs combined with a multilayer perceptron to classify streaks in Asteroid Terrestrial-impact Last Alert System (ATLAS) data. These approaches have focused on classifying small (up to a few hundred pixels wide) sub-images containing streak candidates, but not on extracting streak coordinates directly from large images. An object-detection approach was taken by \citet{kruk2022}, who tested a Google Cloud AutoML pipeline to detect asteroids in \textit{Hubble} images, and \citet{varela2019}, who tested YOLOv2 \citep{redmon2017} to detect satellite streaks in the data of their Wide-Field-of-View (WFoV) System. These object-detection studies have used the classical bounding-box approach, which involves drawing a rectangular box around an object of interest, but the resulting detection and streak-coordinate accuracies suggest that this method may not be optimal for streak detection.

Intending to improve the detection capability beyond the \texttt{StreakDet} pipeline, especially for faint SSO streaks, we built a custom object detection pipeline consisting of a CNN, a recurrent neural network (RNN), and an \texttt{XGBoost} model to detect and return streak coordinates from large images and link objects between observations. Instead of using bounding boxes, our model returns the endpoint coordinates for each detected streak. Our pipeline can be used directly on full CCD images without any separate streak-detection algorithm. The reasoning behind our approach is that if a CNN is better than other methods at recognizing and classifying SSO streaks, as it seems from the results of all the relevant work listed above, then logically, using some other initial method before the CNN is suboptimal. This work tested how high completeness (recall), purity (precision), and coordinate accuracy can be achieved when detecting simulated SSOs using neural networks and determined the apparent magnitude and sky motion ranges of the successful detections.

In the following sections, we start by describing the properties of the simulated \textit{Euclid} VIS images and the training and test sets generated from them. Next, we outline the deep-learning methods utilized. Then, we present and discuss our results obtained with the deep-learning approach. Lastly, we provide our conclusions.

\section{Image data}

\subsection{Simulated \textit{Euclid} VIS images and preprocessing}

\begin{figure*}%
    \centering
    \includegraphics[width=\textwidth]{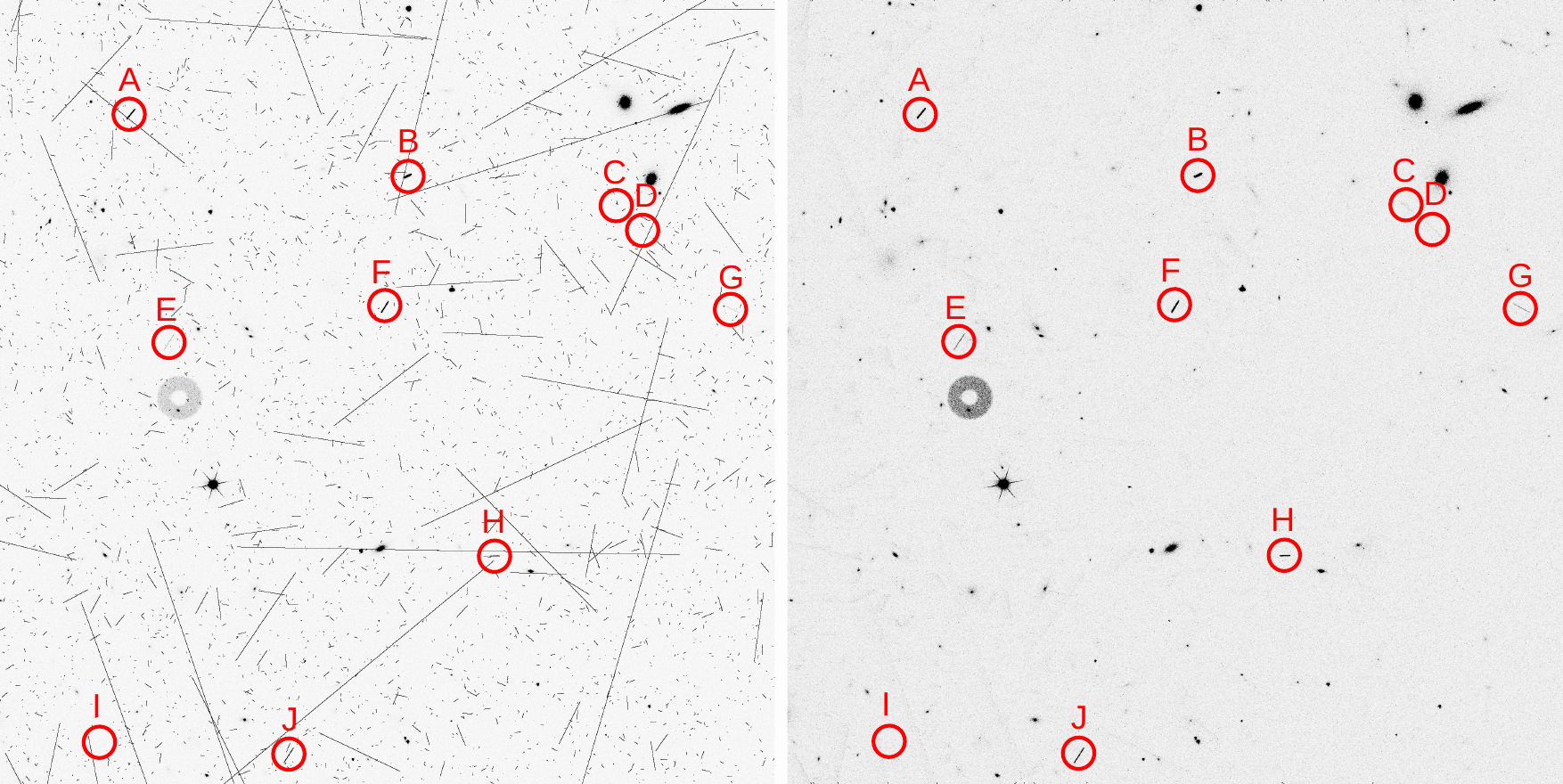}
    \caption{Example of the simulated VIS data. The \emph{left} image shows a quadrant of a raw CCD file with a size of 2048\,$\times$\,2066 pixels (204.8\,$\times$\,206.6 arcseconds). Asteroids are marked with red circles. The other streaks are simulated cosmic rays. On the \emph{right} is the same image after removing cosmic rays. The dark objects above asteroids \emph{C} and \emph{D} are galaxies, the circle below asteroid \emph{E} is a ghost reflection of a bright star, and the spot further down with six outward lines is a star with diffraction spikes.
    Streaks \emph{A}, \emph{B}, \emph{F}, \emph{H}, and \emph{J} are quite bright and easily seen (magnitudes 20.3--22.2), while \emph{C}, \emph{D}, \emph{E}, and \emph{G} are fainter and harder to see (magnitudes 22.6--23.9), and finally, streak \emph{I} is practically invisible (magnitude 25.0). The velocities range from $14\,{\rm arcsec\,h^{-1}}$ (asteroid \emph{B}) to $56\,{\rm arcsec\,h^{-1}}$ (asteroid \emph{D}).}
    \label{fig:Data}
\end{figure*}

To make it easy to directly compare the performance of the deep-learning pipeline to that of \texttt{StreakDet}, we used the same simulated \textit{Euclid} VIS data as \citet{pontinen2020}. The VIS data used by Nucita et al. (in prep.) is also generated with the same software and similar image properties, but it is not identical. We use only the VIS data for asteroid detection because VIS images have longer exposure times and thus reach fainter objects than NISP. The NISP data is valuable for asteroid science, but by using the VIS SSO detections, it should be possible to calculate where the detected SSOs are located in the NISP images without separate asteroid detection processes necessary for the NISP data. We use raw VIS images because they will be quickly available during the \textit{Euclid} mission, which is essential for potential follow-up observations. The following is a summary of the sections in \citet{pontinen2020} and Nucita et al. (in prep.) describing the generation and preprocessing of the simulated VIS image data.

The Euclid Consortium has developed a \texttt{Python} software package named \texttt{ELViS} to generate realistic simulated \textit{Euclid} VIS data (Euclid Collaboration: Serrano et al. in prep.). The simulated images include stars, galaxies, Solar System objects, cosmic rays, and artifacts such as ghosts, diffraction spikes, charge transfer inefficiency (CTI), bleeding, Poisson noise, readout noise, and bias (Fig.~\ref{fig:Data}).

The simulated asteroid population was generated with uniformly distributed random apparent magnitudes between 20 and 26. The apparent velocities range from 1 to $80\,{\rm arcsec\,h^{-1}}$. These magnitude and velocity ranges cause many of the streaks to be practically invisible in the images. Therefore, the essence of the dataset is not that all the streaks could be detected and completeness of 100\% could be achieved, but rather the dataset is built for characterizing the detection envelope, that is, testing where the magnitude and velocity limits of detection are.

Streak angles range from 0\degree{} to 360\degree{} (clockwise from east). The simulated streaks are symmetrical. Thus, within exposures, it is not possible to distinguish between their start and end points. However, the direction becomes apparent when analyzing several exposures. One CCD image contains approximately 25 asteroids on average. The simulated asteroid streaks do not follow realistic rate-of-motion distributions or position-angle distributions but are nevertheless reasonably well-suited for measuring the performance of the pipeline. The brightness of a simulated SSO stays constant within and between exposures. In other words, the potential brightness variation caused by asteroid rotation is not simulated. In reality, some asteroid streaks would have noticeable brightness variation between exposures and even within a single exposure. However, the portion of such asteroids is very small. For asteroids with known rotation periods and brightness variation amplitudes, the median rotation period is 7.2 hours, with 90\% of asteroids spinning slower than once every 2.8 hours, and the median amplitude is 0.35 magnitudes, with 90\% having a smaller amplitude than 0.77 magnitudes \citep{warner2021}.

The size of each CCD is 4096\,$\times$\,4132 pixels, corresponding to approximately 16.9 megapixels. One exposure forms a mosaic of 6\,$\times$\,6 CCDs. The simulated data consist of four dithered exposures, so the total number of CCD images per dataset is 144. The shorter VIS exposures taken alongside the NISP observations were not simulated. We worked with 11 datasets, ten of which focused on faster-moving SSOs (from 10 to $80\,{\rm arcsec\,h^{-1}}$) and one focused on slower ones (from 1 to $20\,{\rm arcsec\,h^{-1}}$). In total, our datasets have 1584 CCD images. The data are in the format of multiextension FITS files, with each quadrant in its own FITS extension. For easier data management and image analysis, we tiled the quadrants into single-extension FITS files before utilizing them in our pipeline.

As a preprocessing step, we removed bright pixels caused by cosmic rays with \texttt{Astro-SCRAPPY} \citep{mccully2018, vandokkum2001}. We used the default parameters of \texttt{Astro-SCRAPPY}, which worked well, and the program removed practically all cosmic rays. After removal, some residuals caused by the CTI effect remain in the images. SSO streaks are not removed in this process due to their different point spread function (PSF) shapes.

Since both the asteroids and the pointing of the telescope move between exposures, the asteroids can move outside the field of view. Therefore, not all asteroids are visible in all four exposures. In our simulated data, the asteroids appear as follows: 37\% in four exposures; 45\% in three exposures; 9\% in two exposures, and 9\% in only one exposure.

\subsection{Training and test sets for deep learning}

\begin{figure}%
    \centering
    \includegraphics[width=\hsize]{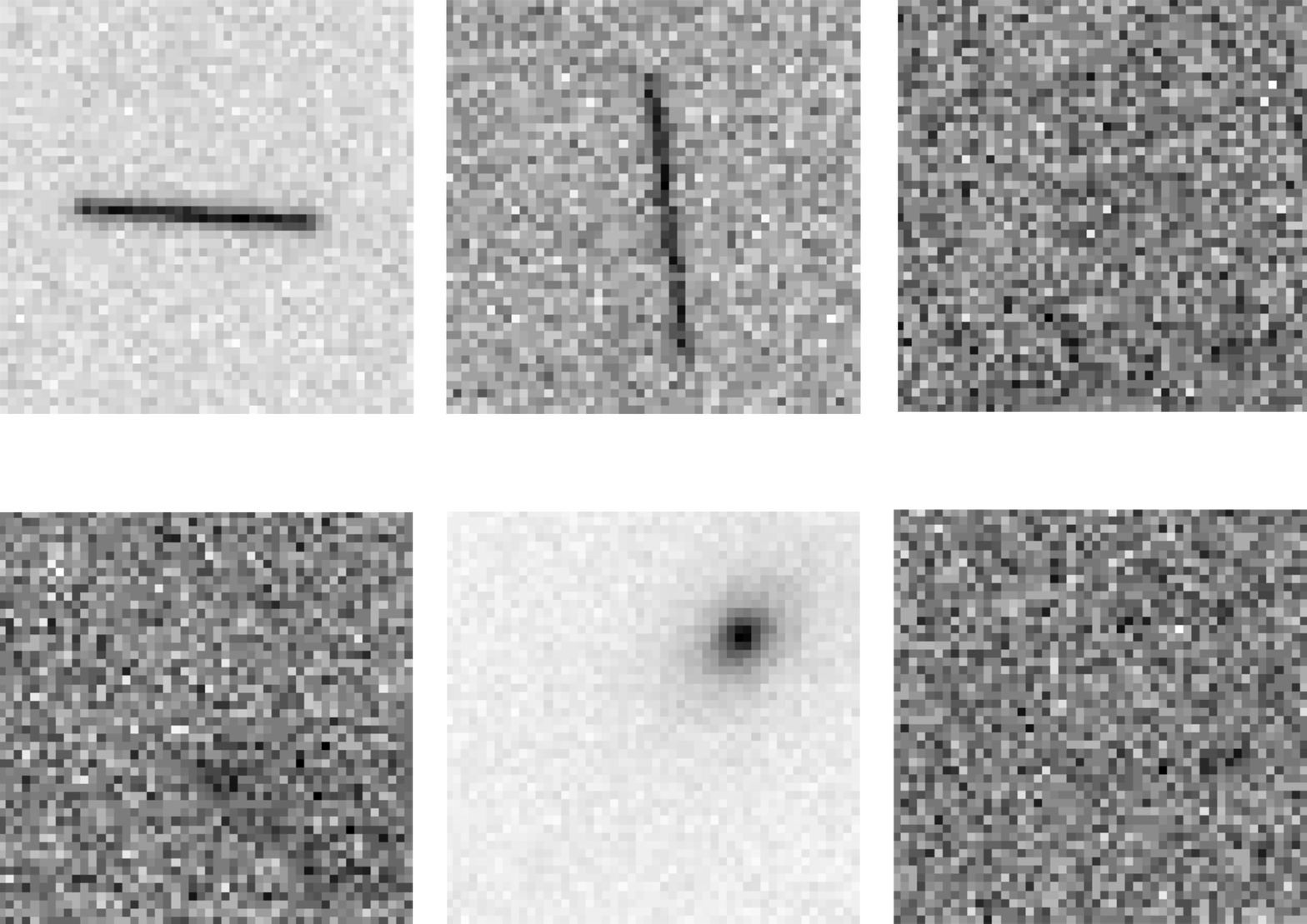}
    \caption{Examples of cutout images generated from the simulated data for the CNN training and test sets. The single-channel cutouts shown are 50 pixels (5\,arcsec) in height and width. The \emph{top} row shows positive training examples, namely asteroid streaks. The streaks are centered training examples. As seen from the third positive example (\emph{top right}), some generated asteroid streaks are so faint that they are practically invisible. The \emph{bottom} row shows negative training examples, specifically images that do not contain asteroid streaks.
    }
    \label{fig:Training_data}
\end{figure}

We developed software to generate different kinds of training and test datasets for the deep-learning models from the simulated \textit{Euclid} images. The basic idea is to extract smaller sub-images or cutouts from the simulated CCDs (Fig.~\ref{fig:Training_data}). The training labels for each training example are the classification label and, for positive examples, also coordinates for the endpoints of the streak visible in the cutout. For consistency during the training and inference stages of the deep-learning pipeline, the leftmost end of a streak is defined as the starting point and the rightmost end as the ending point. Negative examples are cutouts selected randomly from areas of the CCDs not containing asteroids. The size of the cutout images and the ratio between positive and negative training examples can be chosen by the user. Furthermore, either single-channel data (one exposure) or four-channel data (all exposures) can be generated. The training and test sets are saved as CSV files.

Our software can generate either centered images, where an asteroid streak is going through the center of each cutout, or non-centered data, where the streaks are in random positions of the cutouts. The centered data are generated by taking cutouts around the central point of the ground-truth streak. The non-centered data are generated with a sliding-window algorithm so that a window, that is, a test cutout, is first formed from the upper-left corner of a CCD and compared to the ground-truth catalog. If the test cutout overlaps with a full or partial ground-truth streak, it is saved into positive examples and otherwise into negative examples. Then the window moves a given amount of pixels to the right and is again analyzed for ground-truth streaks. This procedure repeats until the whole image is analyzed and divided into positive and negative examples. Since there are many more negative examples than positive ones, especially with small window sizes, only a subset of the negative examples is randomly chosen for the final training or test set. It is also possible to use a filter to determine how far from the edge of the cutout the streak must be visible, in order to avoid positive examples where the streak is barely visible just at the edge of the cutout.

We conducted a series of tests with different training set parameters. Ultimately, we settled on our final configuration, which includes using non-centered cutouts with dimensions of 20 pixels (both height and width), maintaining a ratio of five negative examples to every positive one, and utilizing single-channel data. Additionally, we explored two training approaches: one using data containing streaks of all magnitudes, including practically invisible ones, and another restricted to streaks visible to the naked eye. Both approaches resulted in similar results. With a training set containing invisible positive training examples, the models learned to give negative classifications to streaks that were too faint to detect and give positive classifications only to streaks that were actually visible. Consequently, we chose to incorporate streaks of all magnitudes into the training set, showcasing the models' robustness and adaptability to varying training data compositions.

The pixel values in the FITS files used range from 0 to 65535 ($2^{16} - 1$). However, the vast majority of the image area has pixel values in the range between 200 and 2000, with only the brightest stars (and cosmic rays before their removal) exceeding 2000 and approaching the upper limits of the whole range. The background level is approximately 230, and the analyzed asteroid streaks reside between 230 (the faintest streaks indiscernible from the background) and somewhat over 1000 (streaks from the slowest and brightest objects). Therefore, to lessen the dominance of bright stars and make the pixel value range more useful for asteroid detection, we performed a clipping procedure. Specifically, we set all pixel values below 200 to 200 and values above 2500 to 2500. We settled upon the chosen values after testing different ranges. After clipping, we normalized the pixel value range from 200-2500 to 0-1. This clipping procedure slightly improves the prediction accuracy and speeds up the training of the model.

The test sets are generated from CCD exposures that are not in the training set. We numbered the CCDs in the 6\,$\times$\,6 mosaic starting from one in the lower-left corner and increasing first to the right and then up, ending at 36. We chose CCDs number 11, 18, 22, 25, and 26 of all the mosaics to form the test set and the rest of the CCDs to form the training and validation sets. This test set offers the most uniform distribution of streak lengths and magnitudes and offers sky background from different parts of the mosaics. Approximately 78\% of the streaks are in the training set, 8\% in the validation set, and 14\% in the test set. We combined all the cutouts extracted from the 11 image sets with this logic to create the training, validation, and test sets.

\section{Deep learning}

Deep learning refers to machine learning with deep neural networks \citep{lecun2015}. During the last few years, it has resulted in breakthroughs in many distinct fields, for example, computer vision, natural language processing, machine translation, and game-playing.

Convolutional neural networks (CNN) are a subtype of artificial neural networks, and their distinctive feature is the use of convolutional filters or kernels \citep{lecun1998}. In traditional manually designed computer vision algorithms, the kernels are manually built and tuned, whereas in CNNs the values of the kernels themselves are parameters to be optimized through a learning process from the training data. In addition to the convolutional kernels, CNNs typically also contain pooling layers that combine the outputs of a previous layer by summing or averaging them locally before feeding them as inputs to the next layer. This approach reduces the number of weight parameters in the network, making it possible to build deeper models. Often there are also fully connected layers at the end of the model. As a result of their success, CNNs have been essential in breakthroughs in artificial intelligence, especially in computer vision applications. One subtype of CNNs is the residual neural network (ResNet) architecture, which was developed to enable the use of much deeper neural networks than was previously possible, using skip connections between network layers \citep{he2016}.

Typical computer vision tasks are image classification and object detection. Image classification can refer to binary classification,  which involves detecting whether an image contains a member of a particular object class or not, or to multiclass classification, where the goal is to detect whether an image contains a member of any of several possible object classes. Object detection takes things a step further from classification, and in addition to the classification label, an object-detection model also returns the positions of the detected objects in the image, typically in the form of bounding boxes.

Recurrent neural networks (RNN) are a category of models in which the units, in a sense, loop back into themselves, enabling previous outputs of the model to be used as inputs for the next iteration. This approach is helpful for data such as time series or natural language. One of the main benefits of RNNs is the ability to process inputs of varying lengths. Typical examples of RNN architectures are the Long Short-Term Memory \citep[LSTM;][]{hochreiter1997} and Gated Recurrent Unit \citep[GRU;][]{cho2014} algorithms.

\texttt{XGBoost} (short for eXtreme Gradient Boosting) is an open-source library implementing gradient boosting \citep{chen2016}. Gradient boosting is a machine-learning technique that refers to creating and training an ensemble of simple prediction models, such as decision trees, which would be quite weak when used individually but reach high accuracy when used together as a \enquote{committee} \citep{friedman2000, friedman2001}. \texttt{XGBoost} does not implement neural networks, so using it is technically not deep learning but belongs to a broader group of machine-learning methods. Gradient-boosted trees often outperform neural networks in regression and classification tasks when the input consists of heterogeneous tabular data \citep{shwartz2022, grinsztajn2022}.

\section{Machine-learning pipeline for \textit{Euclid}}

\subsection{Streak detection from raw image data with CNN}

\begin{figure*}%
    \centering
    \includegraphics[width=0.9\hsize]{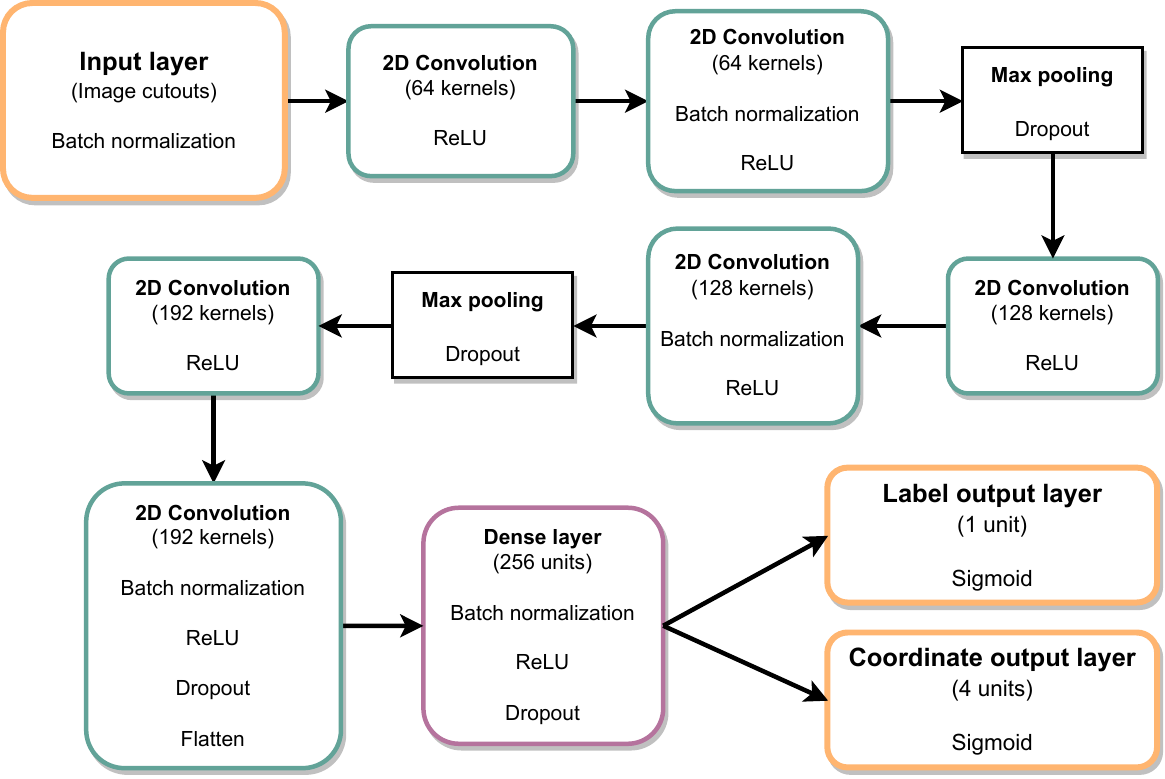}
    \caption{Structure of the CNN model. The model contains 2\,043\,975 trainable parameters. The kernel size of all filters is 3\,$\times$\,3, and the pooling layers have a 2\,$\times$\,2 pool size. The dropout probability for the max-pooling and convolutional layers during training is 0.25, and 0.5 for the dense layer.}
    \label{fig:CNN}
\end{figure*}

We trained a CNN model to classify cutout images into asteroids and non-asteroids. We built the CNN model using \texttt{Tensorflow} \citep{abadi2016} and the \texttt{Keras} API \citep{chollet2015}. To develop the model to be more suitable for streak detection in the manner of object detection, we wrote a custom loss function based on the \texttt{YOLO} object detection model \citep{redmon2016} to both classify the images and return the coordinates for the streak endpoints for positively labeled examples. The loss for the classification label and streak endpoint coordinate accuracy is defined as
\begin{align} \label{eq1}
L ={}& \lambda_{\rm coord} \, \hat{C} \brackets{ \paren{x_1-\hat{x}_1}^2 + \paren{y_1-\hat{y}_1}^2 + \paren{x_2-\hat{x}_2}^2 + \paren{y_2-\hat{y}_2}^2 } \notag\\
&+ \lambda_{\rm noobj} \abs{ C - \hat{C} }^3\,,
\end{align}
where $\lambda_{\rm coord}$ and $\lambda_{\rm noobj}$ are weights, $\hat{C}$ is the ground-truth classification label, which equals 1 when there is a ground-truth object in the image and 0 otherwise, $C$ is the predicted classification label, $x_1$ and $y_1$ are the predicted coordinates of the leftmost end of the streak, $x_2$ and $y_2$ are the predicted coordinates of the rightmost end, and $\hat{x}_1$, $\hat{y}_1$, $\hat{x}_2$, and $\hat{y}_2$ are the ground-truth coordinates. The weights $\lambda_{\rm coord}$ and $\lambda_{\rm noobj}$ are set as 5 and 0.5, respectively, similar to the original YOLO loss function. We tested different weights, but the original values appeared pretty close to optimal. 

When there is no ground-truth streak in the image, the coordinate loss (the first part of the equation) is zero, the coordinate predictions are ignored, and the classification loss (the second part of the equation) is minimized when the classification label approaches the ground-truth label. On the other hand, when there is a ground-truth streak in the image, the coordinate loss is minimized when the predicted coordinates approach the ground-truth coordinates, and the classification loss is again minimized when the classification label approaches the ground-truth label.

Before training and testing the CNN with our custom loss function, we tested training the model for the binary classification task, without coordinate prediction task, with a standard binary cross-entropy loss function. Ultimately, both loss functions achieved practically identical classification accuracies on our datasets.

To optimize the hyperparameters of the CNN model, we ran hyperparameter sweeps using the online service Weights and Biases \citep{wandb}. We tested ResNet models of different depths, as well as simpler CNN models. The sweeps were run on NVIDIA V100 and A100 graphics processing units (GPU) on the high-performance computing platform of the University of Helsinki. In addition, some smaller models were run on a local NVIDIA GeForce RTX 3060 GPU.

After testing, the CNN models worked best for smaller images rather than larger ones. The classification accuracy was fairly uniform for images of varying sizes, but the streak coordinate accuracy was better for small images. Furthermore, using images with four channels (four exposures) did not offer an advantage over images with just one channel (one exposure), but were slower to train. Therefore, we chose to use cutout images with a single channel and widths and heights of 20 pixels. The tested ResNet models did not offer improvements over the simpler models, probably due to the small size of the analyzed images. We settled for a simpler seven-layer CNN model due to its faster runtime compared to the deeper ResNet models. To increase the purity (precision) of the CNN, we chose to use a training set containing five negative training examples for each positive one. Increasing the ratio further did not offer additional advantages but slowed the training. 

Our CNN model uses 2D convolutional layers, batch normalization layers \citep{ioffe2015}, max-pooling layers, and a dense layer (Fig.~\ref{fig:CNN}). The activations in the hidden layers are rectified linear units \citep[ReLU;][]{agarap2018}, and the output units have sigmoid activation functions. To reduce overfitting during the training stage, we used several dropout layers. There is one output unit for predicting the classification label and four for predicting the streak endpoint coordinates (one unit for $x$-start, $y$-start, $x$-end, and $y$-end each). The classification label is a probability, that is, a value between zero and one. Each of the coordinates is a value between zero and one as well, so that the $x$-axis of each analyzed image runs from zero on the left side to one on the right side, while the $y$-axis runs from zero to one from bottom to top. We used the Adam optimizer \citep{kingma2014} and a batch size of 256 for training the model. During training, we used an early stopping callback, monitoring the validation loss with a patience of 30 epochs. Once the validation loss did not improve over 30 training epochs, the model weights corresponding to the epoch with the best validation loss were saved as the final model.

The CNN we trained takes only small cutout images as input, but we need to analyze the whole 4\,k\,$\times$\,4\,k-pixel images of \textit{Euclid}. Therefore, we implemented an algorithm in the style of a sliding window, which involves starting from the upper left corner of the full CCD, analyzing that, then moving a set amount of pixels right, analyzing the small image again, shifting down by the same amount at the end of the row, and so forth, until the whole image is analyzed. In practice, to optimize the runtime of the process, the CCD is divided into small cutouts according to the sliding window positions, and the cutouts are then given as an input to the CNN simultaneously, instead of one at a time, as would be the case with a for-loop implementation. Because the predictions for the full CCDs are made with a sliding-window algorithm, CNNs trained with sliding-window-style training data with non-centered streaks worked better in the analysis of the full CCDs than CNNs trained with centered data.

The outputs, which include the predicted labels and streak coordinates of the sliding-window CNN, are saved into catalogs. The threshold for saving the streak into the catalog is a label prediction higher than 0.5, the default threshold used in classification tasks. We also tested different thresholds, but lowering the threshold decreased purity too much while elevating the threshold decreased completeness.

\subsection{Intra-exposure streaklet merging with RNN}

\begin{figure}%
    \centering
    \includegraphics[width=\hsize]{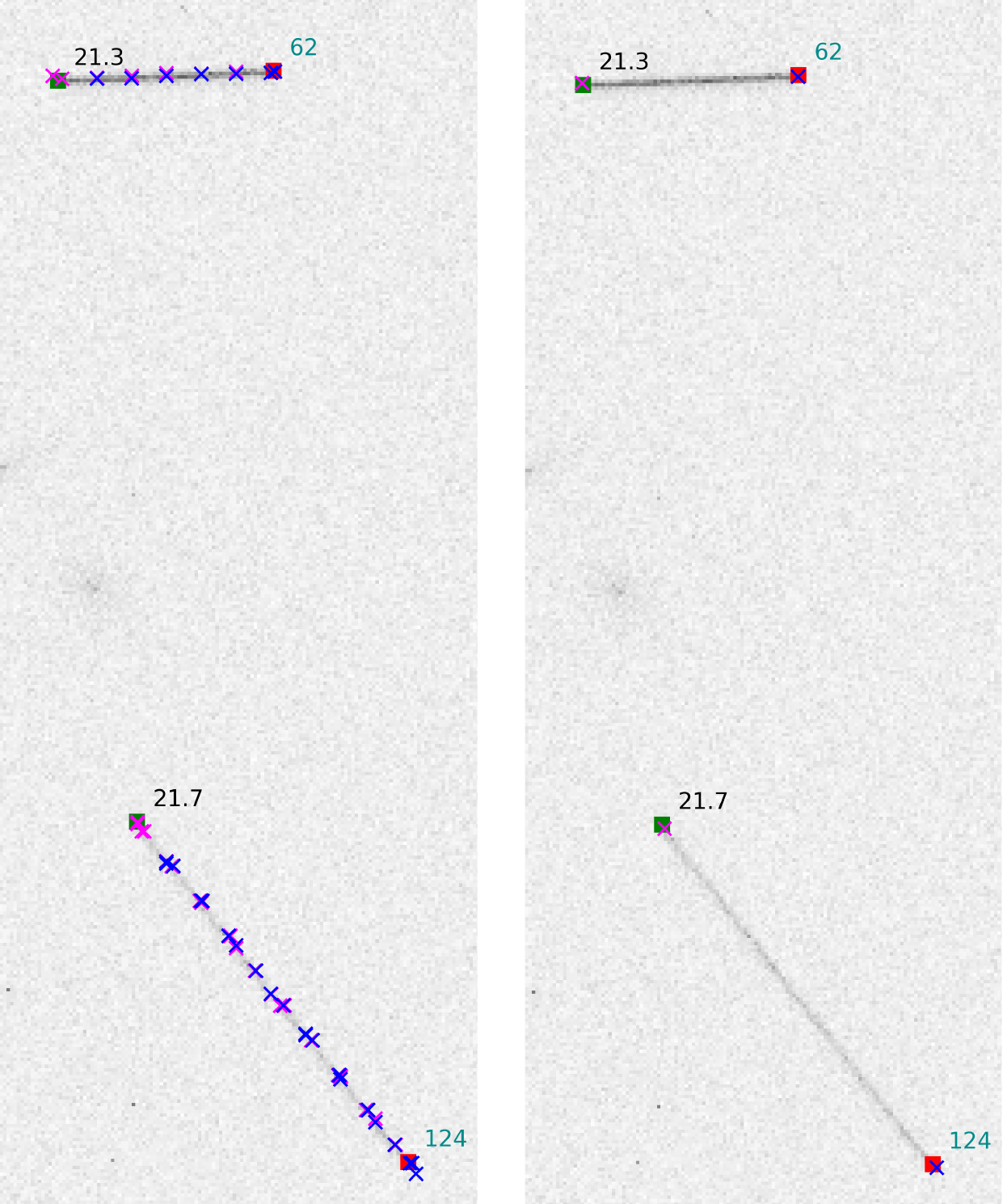}
    \caption{Example of merging streaklet clusters into coherent streaks with RNN. The \emph{left} image shows the detected streaklets before merging, and the \emph{right} image shows the streaks after RNN merging. The images contain two ground-truth streaks whose starting points are marked with green squares and ending points with red squares. The black numbers show the magnitude of the asteroids, and the cyan numbers show the streak lengths in pixels. The magenta X's mark the starting points of CNN detections, and the blue X's mark the ending points. The ground-truth streaks shown here are long, so there are many CNN detection streaklets along them. After merging with RNN, there are single coherent streaks whose coordinates match the ground-truth coordinates quite well.}
    \label{fig:RNN_merge}
\end{figure}

\begin{figure}%
    \centering
    \includegraphics[width=0.5\hsize]{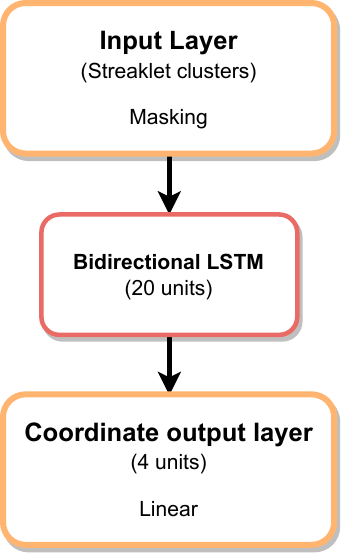}
    \caption{Structure of the RNN model. The model contains 4324 trainable parameters.}
    \label{fig:RNN}
\end{figure}

Our CNN model analyzes 20-pixel wide cutout images, yet many asteroid streaks are longer than that and appear in multiple cutouts as streaklets, which are shorter snippets of full streaks. We use an RNN model, a bidirectional LSTM, to merge these streaklets into full streaks (Fig.~\ref{fig:RNN_merge}). The model has an input layer, a masking layer for managing variable-sized inputs, a bidirectional LSTM layer with 20 units, and an output layer with four units with linear activation functions (Fig.~\ref{fig:RNN}). We use mean absolute error as the loss function and NAdam \citep{dozat2016} as the optimizer. We use a batch size of 512 and an early stopping callback with patience of 500 epochs on the validation loss. Similar to the CNN model, we built the RNN model with \texttt{Tensorflow} and \texttt{Keras}. According to our tests, the two-stage process of using the CNN on small windows followed by the RNN merging process results in better coordinate accuracy than using the CNN directly on larger windows and omitting the RNN stage.

For generating a training set, we developed an algorithm that groups nearby and similarly angled streaklets in the CNN output catalogs that match the coordinates of ground-truth streaks. These groups, generated from the training and validation set CCDs, form the training and validation set for the RNN. In other words, the predicted coordinates and CNN output labels of groups of streaklets approximately at the position of ground-truth streaks, that is, true-positive CNN predictions, are the training data (\vec{X}) for the RNN, while the coordinates of the corresponding ground-truth streaks' endpoints are the labels to be learned (\vec{Y}). The RNN is then trained with these data. The trained RNN can merge streaklet clusters into full streaks when applied to the sliding-window CNN catalogs.

Before generating the training, validation, and test sets, we remove so-called static streaklets. Streaks that stay in the same location in multiple exposures are static, meaning they are not SSOs but objects that stay stationary on the sky, such as galaxies. Therefore, we can remove them from the catalogs. Approximately 17\% of all streaklets in our pipeline were static and therefore removed at this stage.

The streaklet clusters are located all over the image mosaic, so to make training the RNN model easier, each cluster is moved closer to the origin of the pixel coordinate system for the duration of the RNN training and prediction. Specifically, the clusters are moved to the 0--250 pixel coordinate range along both $x$ and $y$-axes, and then normalized to between $-1$ and 1. After the model has merged the streaks, the coordinates of the resulting streak are denormalized and moved back to the original coordinates of the streaklet cluster. The same coordinate shift is also done in the testing phase. Including the CNN output classification labels (i.e., probabilities) in the RNN input data improves the RNN coordinate prediction accuracy by approximately 10\%, compared to inputting only the CNN output coordinates. This is probably due to CNN labels helping RNN to recognize outliers, such as false-positive streaklets whose prediction label differs from those in the rest of the streaklet cluster.

During the testing stage, streaklet clusters are created from the CNN output catalogs of the test set CCDs. Nearby streaklets located approximately along the same line are grouped, and the groups are analyzed with graph theory so that each streaklet belongs to only one group, similar to the training stage, except omitting the ground-truth coordinates. The streaklet groups are then fed to the RNN, which merges the clusters into coherent streaks.

\subsection{Inter-exposure streak linking with \texttt{XGBoost}}

\textit{Euclid} observes the same position of the sky multiple times in a row. Stars and galaxies appear in the same sky coordinates in each exposure, but SSOs do not because they move. The motion is typically used for distinguishing SSOs from other objects in the sky. For \textit{Euclid}, we can use the movement of SSOs by linking streaks that appear approximately along the same line in multiple exposures into so-called multi-streaks. In other words, a multi-streak consists of several streaks (from two to four) caused by the same asteroid, appearing along the same line in separate exposures.

We tested a non-machine-learning streak-linking algorithm adapted from \citet{pontinen2020}, an RNN model, and an \texttt{XGBoost} model to do the linking. Both machine-learning models outperformed the non-machine-learning algorithm, and the \texttt{XGBoost} model further outperformed the RNN model, so we settled on using the \texttt{XGBoost} model in the final pipeline.

Before running the streak-linking algorithms, we apply another iteration of the static-streak removal, now to the streaks merged with RNN. Almost all static streaklets are removed already after the sliding-window CNN stage, but a small number of full streaks (0.11\% out of all streaks) can only be removed after the merging with RNN.

To link streaks by the same SSO in multiple exposures, we first tested the so-called multi-streak analysis, which was used in \citet{pontinen2020} to decrease the number of false-positive streaks. We adapted the algorithm to use the CNN+RNN catalog data and focused on the following parameters: minimum number of streaks in a multi-streak combination; maximum length difference between streaks; maximum and minimum distance between streaks compared to their lengths; maximum angle difference between streaks; maximum angle difference between the single streaks and the common multi-streak line; and maximum difference in median pixel flux values of the streaks.

As alternative machine-learning versions of the streak-linking process, we tested RNN and \texttt{XGBoost} models. We generated permutations of possible multi-streak candidates and trained the models to classify them into asteroids and non-asteroids. For training, the permutations are generated from the detected streaks in the CNN+RNN output catalogs of the training set CCD exposures. The multi-streak candidates consist of two, three, or four streaks within 6\degree{} of their common line, all of which are from different exposures. The members of the multi-streaks are compared to ground-truth streaks, and multi-streak candidates whose members all match ground-truth streaks receive a positive ground-truth label. Multi-streak candidates with at least one false-positive member receive a negative ground-truth label.

Before training and predicting, we move the multi-streaks closer to the origin of the pixel coordinate system, similar to the streaklet merging with RNN. Then, to give as much information as possible to the classifier models, we add line solution parameters $m$ and $c$ (as in $y = mx + c$) to each streak in the data, as well as a few differently calculated pixel fluxes. We also include the angles and distances between each streak in a multi-streak. After all the steps, each streak in a multi-streak has 25 numerical features, including exposure, CNN classification labels (min, max, median, mean), coordinates, angle, length, pixel flux (min, max, median, mean, total, total divided by length), line solution parameters $m$ and $c$, and angles and distances to other streaks belonging to the same multi-streak. The only difference between the data for the streak-linking RNN and \texttt{XGBoost} models is that \texttt{XGBoost} accepts only a fixed-size array as input. For RNN, each input data example consists of between two and four arrays, each with 25 features. For \texttt{XGBoost}, we flattened the four arrays into single arrays with 100 features. We filled the missing features with NaN values for cases with fewer than four streaks.

The tested RNN model for streak linking has almost the same model architecture as the RNN model for streaklet merging. It has an input layer, a masking layer for managing variable-sized inputs, and a bidirectional LSTM layer with 20 units. The main difference is that it has only one output unit using a sigmoid activation function for binary classification. Also, the loss function is binary cross-entropy.

To optimize the \texttt{XGBoost} hyperparameters, we ran the RandomizedSearchCV function of \texttt{scikit-learn} \citep{scikit-learn} over eight hyperparameters with five-fold cross-validation. After optimization, the best model consisted of 644 gradient-boosted trees, each with a maximum depth of 11.

Because all permutations of multi-streak candidates are generated, many duplicates remain after the classification. For example, if the ground-truth multi-streak contains members in all four exposures, there are different permutations of it containing members in only two or three exposures. Therefore, after the \texttt{XGBoost} classification, the program checks if all members of some positively classified multi-streak are included in another positive multi-streak having members in more exposures. If so, the multi-streak with fewer members is removed.

\subsection{Training and inference pipelines}

\begin{figure*}%
    \centering
    \includegraphics[width=\hsize]{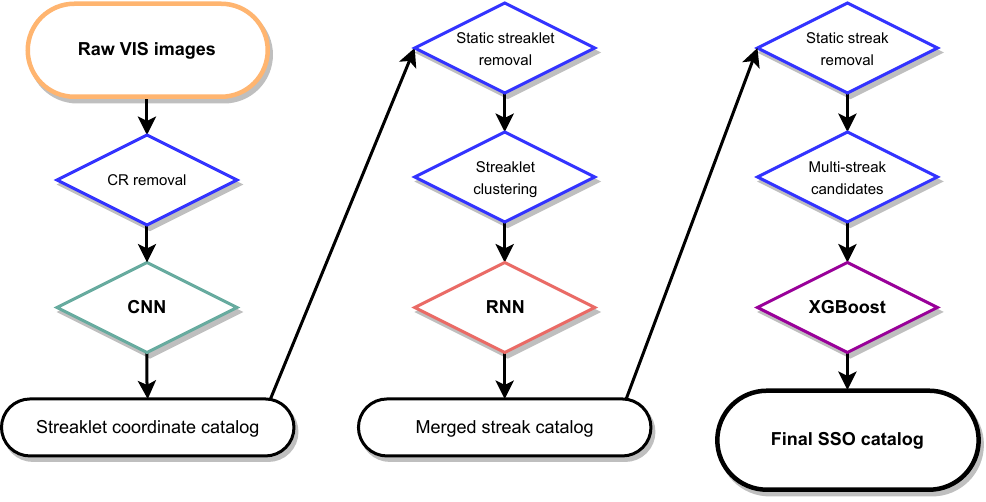}
    \caption{Whole inference pipeline visualized.}
    \label{fig:Pipeline}
\end{figure*}

Combining all the previously mentioned parts, the training of the pipeline consists of the following stages. %

\begin{enumerate}
\item Generating a training set for the CNN from full simulated \textit{Euclid} CCD exposures.
\item Training the CNN with the training data.
\item Running the CNN through the training images with the sliding-window algorithm and generating catalogs of the detected streaklets.
\item Removing static streaklets from the catalogs.
\item Generating a training set for the RNN to merge the streaklets within exposures.
\item Training the RNN with the training data.
\item Using the RNN to merge the streaklet clusters of the CNN output catalogs into unified streaks.
\item Removing static streaks.
\item Generating a training set for the \texttt{XGBoost} model from multi-streak candidates.
\item Training the \texttt{XGBoost} model with the training data.
\end{enumerate}

The training of the CNN takes a few hours on average, depending on the exact hyperparameters used. The training of the RNN typically takes a few tens of minutes, as it contains fewer parameters than the CNN. The training of the \texttt{XGBoost} model is fast and happens in the order of minutes.

On the other hand, the inference pipeline with trained models (Fig. \ref{fig:Pipeline}) is as follows.

\begin{enumerate}
\item Running the CNN through the images with the sliding-window algorithm and generating catalogs of the detected streaklets.
\item Removing static streaklets.
\item Using the RNN to merge the streaklet clusters of the CNN output catalogs into unified streaks.
\item Removing static streaks.
\item Using the \texttt{XGBoost} model to classify multi-streak candidates into asteroids and non-asteroids.
\end{enumerate}

The inference with trained models is relatively fast. The exact times depend on the hardware used and the number of streaks in the images. With an NVIDIA GeForce RTX 3060 GPU, the CNN analyzes a typical CCD with the sliding-window algorithm in five seconds and a complete observation (a field) of 144 CCDs (four exposures of 36 CCDs) in approximately 12 minutes. The streaklet merging with RNN takes approximately six minutes for a field, including the time for clustering nearby streaklets, which takes almost all of the time used in this step. The inter-exposure streak linking takes approximately four minutes for a complete observation, most of which is used again by a non-machine-learning step, namely generating the multi-streak candidates. The classification of multi-streak candidates with \texttt{XGBoost} takes just a few seconds for a field.

Removing static streaklets takes approximately 20 minutes per complete observation. The code for static-streak removal is written in standard \texttt{Python}, making it slow, and leaving room for optimization. Removing static merged streaks is somewhat faster because there are fewer streaks to be analyzed, and it takes approximately four minutes. All in all, the inference with the whole pipeline takes approximately 46 minutes for a typical field of 144 CCDs. The total time it takes \textit{Euclid} to observe a field is approximately 70 minutes, so theoretically, the deep-learning pipeline could keep up with just one GPU and one central processing unit (CPU). Furthermore, there is much room for optimization in the non-machine-learning parts of the pipeline, which currently take most of the running time.

Our deep-learning pipeline outperforms \texttt{StreakDet} in terms of speed, especially when parallelization is not considered. Running a single instance of \texttt{StreakDet} on a field of 144 CCDs takes over two hours, not including post-processing steps. The non-machine-learning processes in the deep-learning pipeline have much room for optimization to achieve even faster execution. Conversely, optimizing the runtime of \texttt{StreakDet} is more challenging as it is already written in well-optimized C++ code. It is possible to parallelize \texttt{StreakDet} by running multiple instances of the software simultaneously on different CPUs, with each instance analyzing different CCD exposures, which reduces the total running time. Comparably, the deep-learning pipeline can be parallelized easily with \texttt{Tensorflow} to utilize multiple GPUs.

\section{Results}

We evaluated the performance of our pipeline by using accuracy, completeness, and purity as metrics. In machine-learning literature, recall is a more commonly used term for completeness, and precision is used instead of purity. In other contexts, completeness may be referred to as sensitivity, hit rate, or true-positive rate, while purity may be referred to as positive predictive value.

For classification tasks, accuracy is defined as 
\be
Accuracy = \frac{\rm TP + TN}{\rm TP + FN + TN + FP} = \frac{\rm TP + TN}{\rm P + N} \;,
\ee
where $\rm TP$ stands for the number of true-positive predictions, $\rm TN$ represents the number of true-negative predictions, $\rm FN$ refers to the number of false-negative predictions, and $\rm FP$ indicates the number of false-positive predictions. The quantity $\rm P$ is the number of actual positive cases in the data, which can be calculated as $\rm TP + FN$, and $\rm N$ is the number of actual negative cases, which is $\rm TN + FP$. In essence, accuracy measures what fraction of all predictions are correct.

Similarly, completeness is defined as
\be
Completeness = \frac{\rm TP}{\rm TP + FN} = \frac{\rm TP}{\rm P} \;.
\ee
In other words, completeness shows what fraction of actual positive cases are predicted to be positive.

Furthermore, purity is defined as
\be
Purity = \frac{\rm TP}{\rm TP + FP} = \frac{\rm TP}{\rm PP} \;,
\ee
where $\rm PP$ is the total number of positive predictions. In other words, purity measures what fraction of all positive predictions are true positives.

\subsection{Results of streak detection with CNN}

\begin{figure*}%
    \centering
    \includegraphics[width=\textwidth]{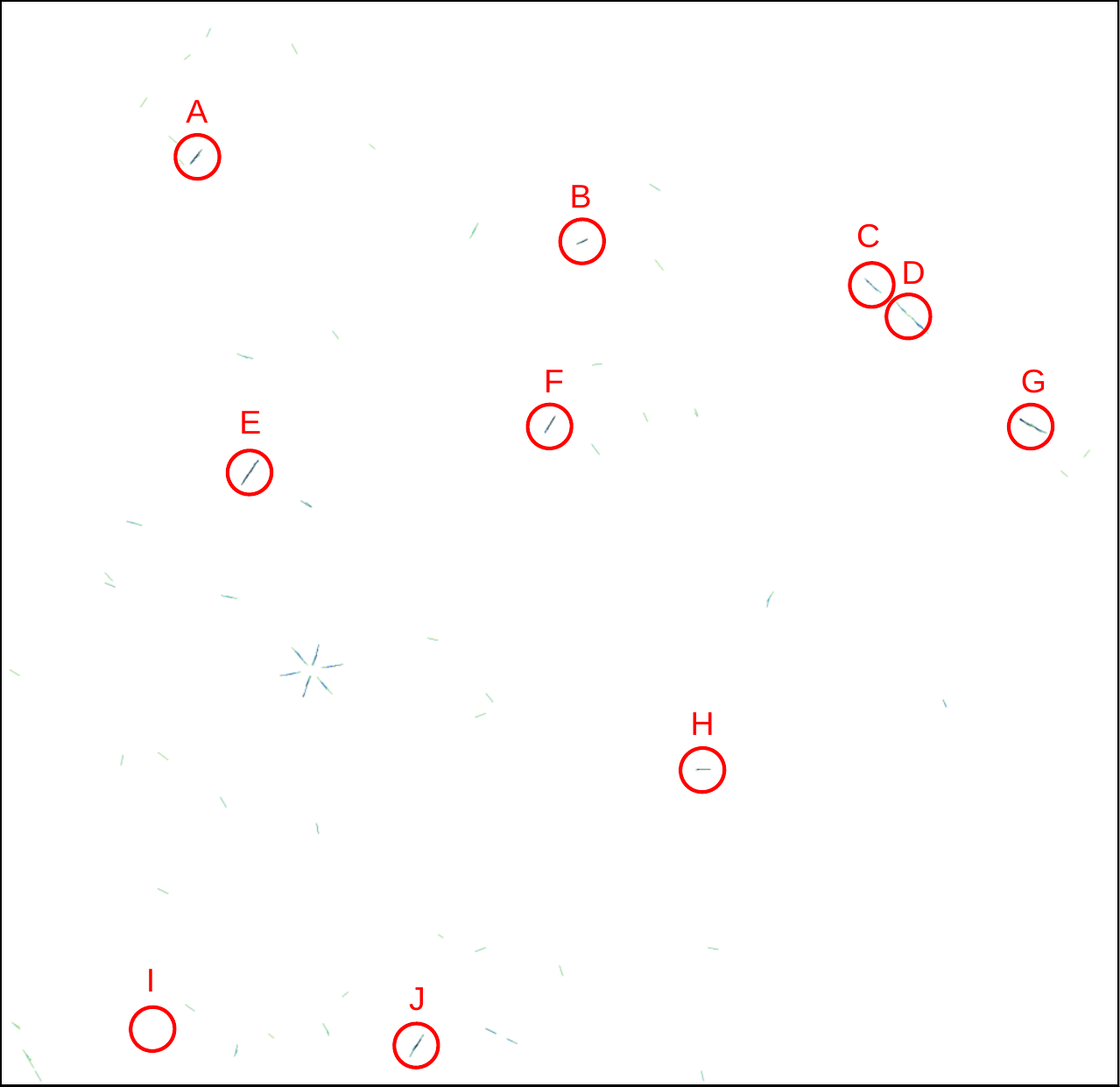}
    \caption{Output catalog of the sliding-window CNN visualized. The input image given to the CNN was the same as in Fig.~\ref{fig:Data}. The detected streaklets in the CNN output catalog are drawn in their predicted coordinates, and the pixel brightness corresponds to the prediction label certainty of the CNN, i.e., the darkest pixels have the highest predicted labels. Streaks with predicted labels of smaller than 0.5 are left out. Pixels with overlapping streaks have been set to the highest predicted label value between those streaks. In this example, the CNN method has detected all asteroids except for the practically invisible asteroid \emph{I}. There are multiple false-positive detections, although their prediction labels typically have lower values than the asteroid streaks. The six streaks caused by a bright star below asteroid \emph{E} are notable exceptions.}
    \label{fig:CNN_output}
\end{figure*}

\begin{figure*}%
    \centering
    \includegraphics[width=\textwidth]{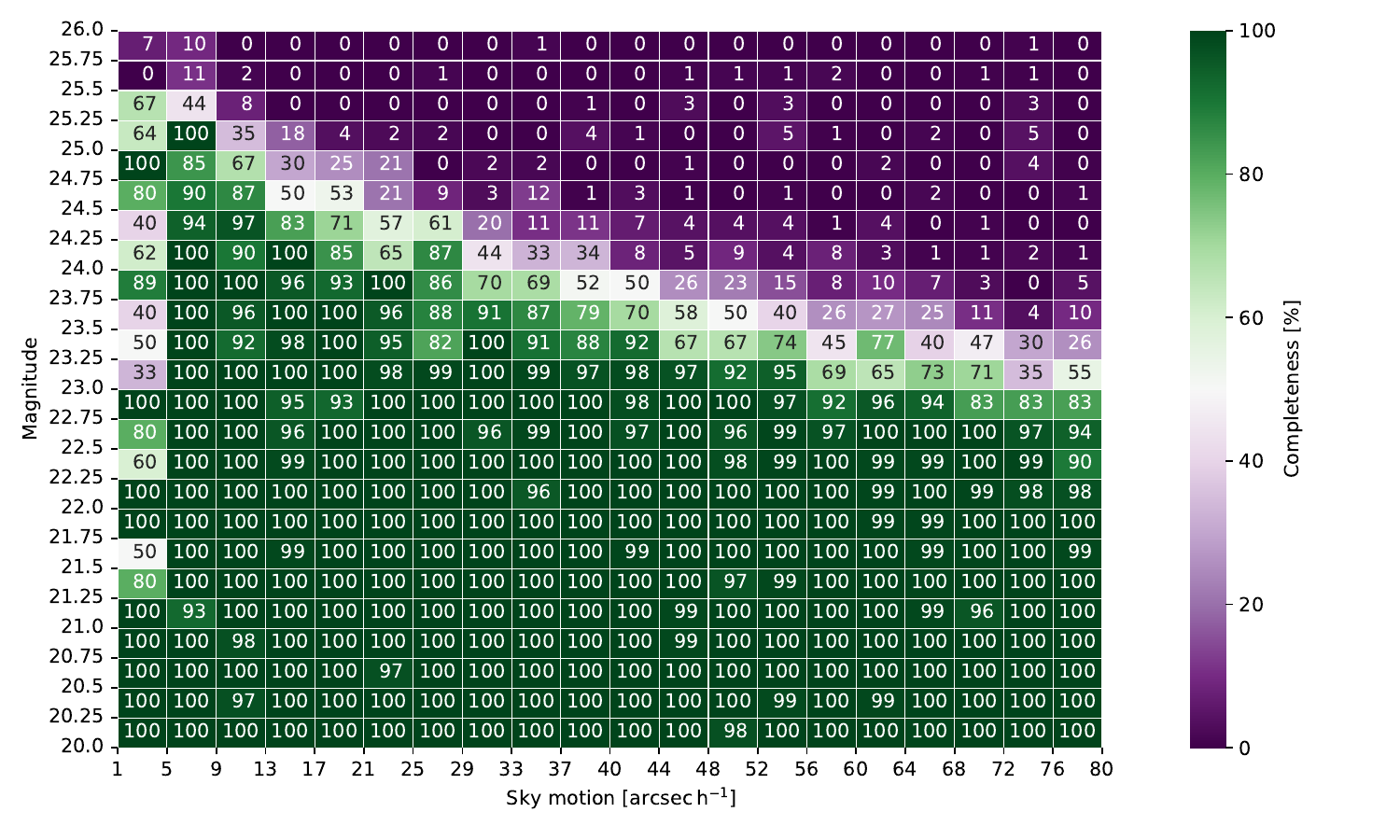}
    \caption{CNN cutout test set classification completeness as functions of apparent magnitude and apparent motion. The values on both axes mark the bin edges. For example, the bin between sky motions of 9 and 13$\,{\rm arcsec\,h^{-1}}$ and magnitudes of 20.25 and 20.5 shows that the detection completeness is 97\% for streaks created by all simulated SSOs between those values. There are 91 ground-truth streaks per bin on average.}
    \label{fig:Heatmap_CNN}
\end{figure*}

Table~\ref{table:cnntable} displays the classification and coordinate prediction results for the CNN when applied to the cutout training, validation, and test sets, that is, before running the sliding-window analysis. The achieved accuracy, purity, and completeness stay fairly consistent for all sets, showing that there is no notable overfitting to the training set. The completeness of approximately 60\% appears low, but as the dataset is built for characterizing the detection limits, it contains many extremely faint, and thus practically invisible, simulated asteroid streaks. Furthermore, non-centered training and test sets generated with the sliding-window algorithm are more challenging than centered sets. Using training and test sets consisting of centered streaks achieves better metrics on the centered sets but falls short on non-centered sets and the sliding-window analysis.

The CNN reaches a purity of 97\% for the cutout training and test sets. However, when applying the CNN to full images with the sliding-window algorithm, the number of false-positive detections rises because a vast majority of the windows are negative. The false positives are typically caused by remnants of removed cosmic rays, which can resemble faint asteroid streaks. Also, the CNN sometimes confuses diffraction spikes from bright stars with bright asteroid streaks (Fig.~\ref{fig:CNN_output}).

The limiting magnitude of detection depends on the object's sky motion (Fig.~\ref{fig:Heatmap_CNN}). The reason is that a fast-moving object's flux is divided among a larger number of pixels than the flux of a slow-moving object, making fast objects appear fainter and thus harder to detect. On the other hand, the slowest-moving objects in our dataset resemble point sources, making it difficult to differentiate them from stars and resulting in lower completeness compared to faster-moving objects.

\begin{table} 
\caption{CNN results for non-centered cutout training, validation, and test sets.}
\label{table:cnntable}
\centering
\begin{tabular}{lccc}
\hline\hline
Parameter &Training &Validation &Test \\
\hline
Accuracy &93.3\% &93.2\% &92.9\% \\
Purity (Precision) &97.6\% &97.7\% &97.0\% \\
Completeness (Recall) &61.4\% &60.9\% &59.3\% \\
True positives &145933 &16094 &26034 \\
True negatives &1185564 &131738 &218531 \\
False positives &3536 &372 &794 \\
False negatives &91873 &10342 &17831 \\
Median coordinate error &0.135 &0.134 &0.138 \\
Mean coordinate error &0.376 &0.384 &0.401 \\
\hline
\hline
\end{tabular}
\tablefoot{
 The training, validation, and test sets contain five times more negative training examples than positive ones. The units for coordinate errors are pixels.}
\end{table}

\subsection{Results of intra-exposure streaklet merging with RNN}

\begin{figure*}%
    \centering
    \includegraphics[width=\textwidth]{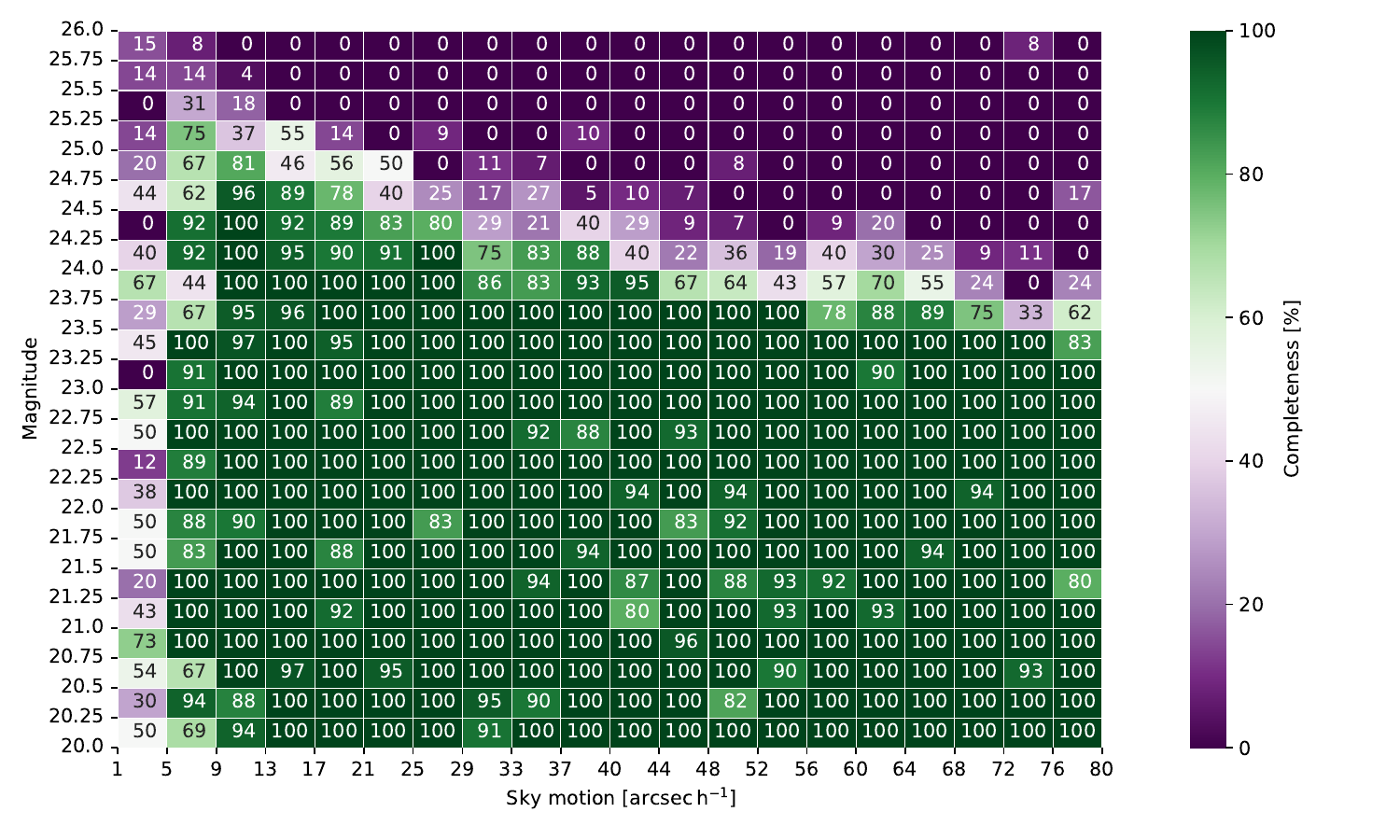}
    \caption{CNN+RNN pipeline detection percentage as functions of apparent magnitude and apparent motion. The values on both axes mark the bin edges. There are 12 ground-truth streaks per bin on average.}
    \label{fig:Heatmap_mode2}
\end{figure*}

After the sliding-window CNN and RNN merging steps, the achieved total completeness is 68.5\% (3956 true positives over 5774 ground-truth streaks). The value is higher than the CNN classification completeness during the cutout training and test phase. This can be explained by the fact that during the cutout analysis, long streaks are divided over multiple cutouts, and in some cutouts, they are more difficult to detect than others, lowering the total cutout classification completeness. However, during the sliding-window CNN and RNN stages, not every single streaklet is needed for the streak to be detected as a whole. This effect is visible in Fig.~\ref{fig:Heatmap_mode2}, where the limiting magnitude for fast-moving objects is notably fainter than during the cutout classification phase shown in Fig.~\ref{fig:Heatmap_CNN}. The achieved completeness appears to be relatively close to the practical limit with the dataset used, judging by how faint different streaks appear in the images.

The purity at this point is only 6.4\% (3956 true positives and 345 duplicates out of 66798 total detections). In other words, there are over 15 times more false positives than true positives. The sharp decrease in purity from the CNN cutout training and test phases is explained by the fact that, during the sliding-window analysis, there is a much larger number of negative windows compared to positive ones, and therefore to achieve high purity during the sliding-window analysis, the achieved purity during the cutout CNN training stage should be practically 100\%. However, focusing on achieving extremely high purity during training inevitably diminishes completeness, typically quite drastically. Since the inter-exposure streak-linking stage can very efficiently improve purity, focusing on purity at the cost of completeness during the intra-exposure stage would be counterproductive.

\begin{figure*}%
    \centering
    \includegraphics[width=\textwidth]{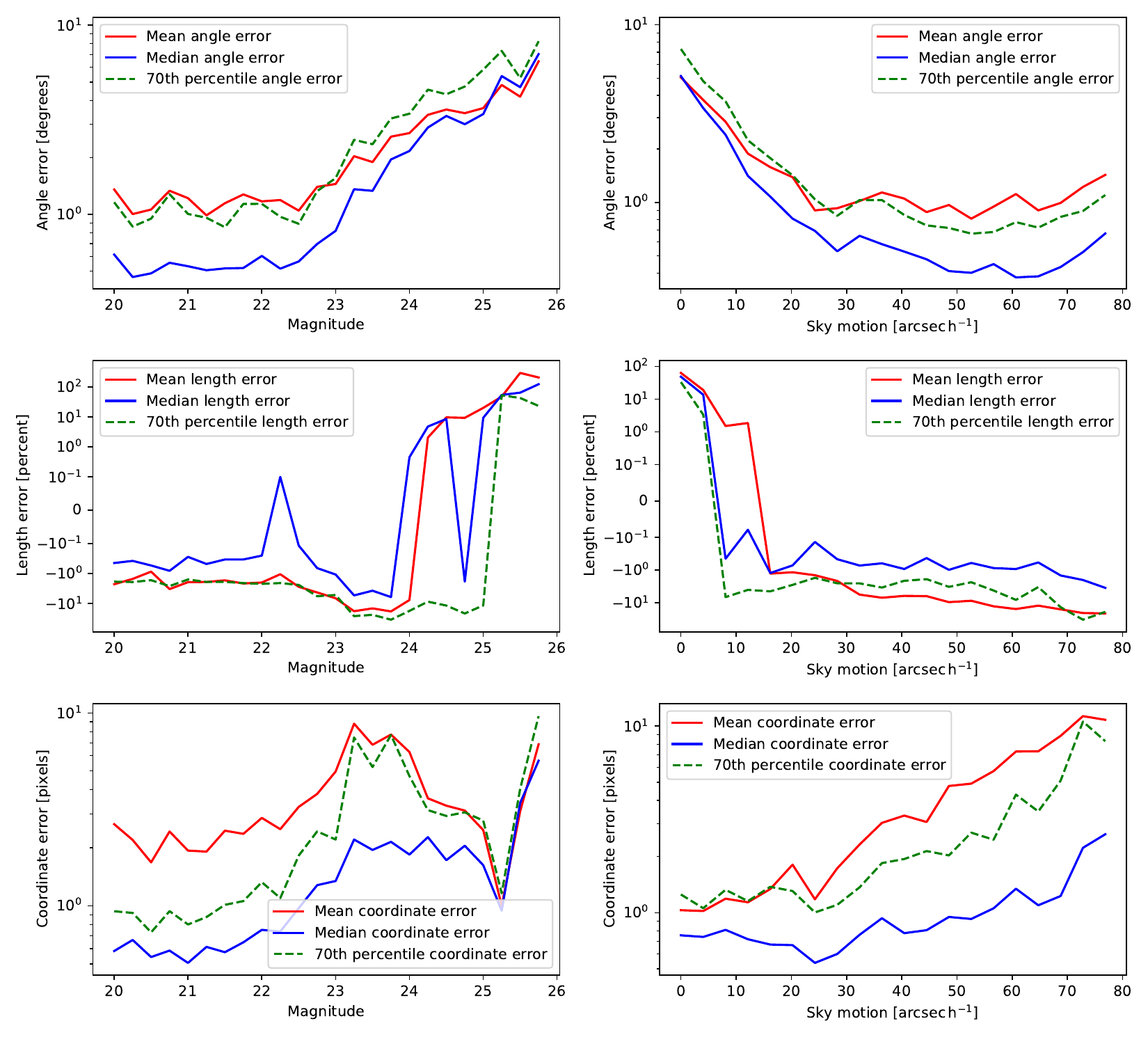}
    \caption{CNN+RNN detection errors for angles, lengths, and coordinates as a function of magnitude and length. The $y$-axes of the \emph{top} and \emph{bottom} rows are logarithmic, while the $y$-axes of the \emph{middle} row are in symmetric logarithmic scale with values between $-10^{-1}$ and $10^{-1}$ in linear scale. In the \emph{middle} row, negative $y$-values indicate that the detected lengths are shorter than the ground-truth lengths, while positive values indicate that the detected lengths are longer than the ground truth. The 70th percentile length errors in the \emph{middle} plots are technically 30th percentile since the length errors are typically negative.}
    \label{fig:Errors}
\end{figure*}

The accuracy of the streak angle is higher for bright streaks than for faint ones (Fig.~\ref{fig:Errors}). For example, in the magnitude range of 20--21, containing the brightest analyzed streaks, the average angle error is \ang{1.1}, and the median angle error is \ang{0.5}. The angle errors remain more or less constant until magnitude 23, after which they increase rapidly. For all magnitudes, the average angle error is \ang{1.6}, and the median angle error is \ang{0.7}. Generally, the angle accuracy improves with streak length because the angle is calculated from the streak endpoint coordinates, and the endpoint coordinate accuracy has a smaller effect on the streak angle accuracy as the length of the streak increases.

For streaks brighter than 21 magnitudes, the average length error is $-2.4$\%, whereas the median length error is $-0.5$\%. A negative sign means that the streaks detected are shorter than the ground truth, whereas a positive sign means that they are longer. As before, the length errors are quite consistent up to magnitude 24, after which they increase and change from negative to positive, indicating that the streaks are estimated to be longer than their actual length. When looking at length errors as a function of apparent motion, streaks corresponding to slower motions are estimated to be too long, whereas streaks corresponding to faster motions are estimated to be too short. The faintest detected streaks are also the shortest, which explains the change from negative to positive errors for the faintest magnitudes.

The coordinate error is defined as the difference between the ground-truth middle-point coordinates of a streak and the estimated middle-point coordinates. For streaks with magnitudes between 20 and 21, the average coordinate error is 2.4 pixels, corresponding to 240 milliarcseconds, and the median coordinate error is 0.6 pixels or 60 milliarcseconds. This error increases for both faint and long streaks. For streaks of all magnitudes, the average coordinate error is 3.8 pixels, or 380$\,{\rm mas}$, and the median coordinate error is 0.8 pixels, or 80$\,{\rm mas}$.

A substantial part of the coordinate error is the length error, because when a streak's length is estimated inaccurately, the middle point of the streak is typically also inaccurate. Analyzing only streaks with length errors of less than 0.1 pixels, corresponding to 4.3\% of all detected streaks, the average coordinate error for all magnitudes decreases to 0.7 pixels (70$\,{\rm mas}$) and the median error to 0.6 pixels (60$\,{\rm mas}$).

\subsection{Results of inter-exposure streak linking with \texttt{XGBoost}}

\begin{figure*}%
    \centering
    \includegraphics[width=\textwidth]{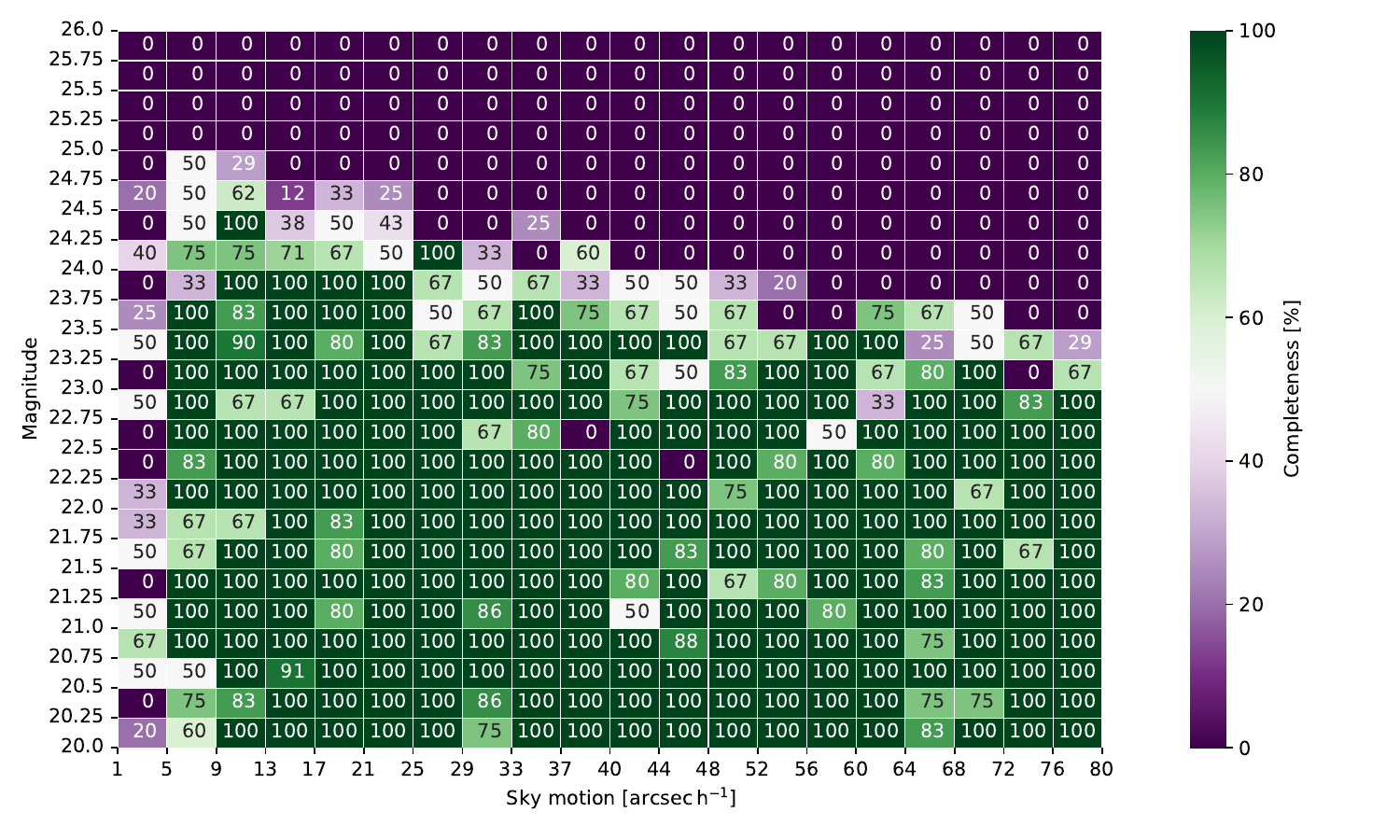}
    \caption{CNN+RNN+\texttt{XGBoost} detection completeness of multi-streaks as functions of apparent magnitude and apparent motion. There are 3.6 ground-truth multi-streaks per bin on average.}
    \label{fig:HeatmapMS}
\end{figure*}

Due to less accurate streak coordinates and angles, the adapted non-machine-learning algorithm did not work as well for linking streaks between exposures as it did for \texttt{StreakDet}'s results in \citet{pontinen2020}. Because of lower coordinate accuracy, strict filtering parameters aiming at high purity resulted in low completeness, while using a broader parameter space resulted in higher completeness but lower purity. Either way, the levels were below the results of the \texttt{StreakDet} pipeline.

The RNN worked much better than the non-machine-learning algorithm and achieved higher completeness and purity. \texttt{XGBoost} further improved completeness, slightly above RNN's level, and it also offered a relatively substantial boost in purity compared to both the non-machine-learning algorithm and RNN. The achieved completeness of \texttt{XGBoost} was 59.1\% (Tables~\ref{table:mstable1}~and~\ref{table:mstable2}), compared to 58.9\% achieved by the inter-exposure RNN, when counting only so-called clean hits, which refer to multi-streaks, whose all constituent streaks are true positives.

There is also a small number of so-called semi-hits, which can be divided into strong and weak ones. Strong semi-hits are multi-streaks containing at least two true-positive streaks but also one or two false-positive streaks. Weak semi-hits contain only one true-positive streak and from one to three false-positive parts. The \texttt{XGBoost} model achieved higher completeness values than the RNN also when accepting strong and/or weak semi-hits. Most of the duplicate detections are caused by the fact that all possible multi-streak permutations are generated before the classification task, and some different permutations still exist after the simple duplicate removal. A more sophisticated algorithm for joining the permutations should be able to decrease the number of duplicates. 

The purity reached by \texttt{XGBoost} was 95.4\%, compared to 85.9\% for the RNN. The miss rate or false negative rate (the complementary percentage of completeness) of the \texttt{XGBoost} model was 37.9\%. The misses consist mainly of very faint or invisible streaks, most of which were missed already by the previous stages of the pipeline. In cases where objects face detection challenges during the previous intra-exposure merging stage, that is, some streaks from these objects remain undetected, the subsequent linking step becomes more challenging or impossible. This leads to a reduction in completeness within these specific bins.

Figure~\ref{fig:HeatmapMS} shows the completeness heatmap of the entire pipeline, encompassing the CNN, RNN, and \texttt{XGBoost}, with a similar logic to Fig.~\ref{fig:Heatmap_mode2}. The plot contains only the clean hits, and the integrated completeness over all bins is 59.1\%. Here the completeness is calculated for asteroids that appear in at least two exposures. Objects appearing in only one exposure cannot be linked and are therefore lost during this stage. In our dataset, approximately 9\% of objects appear in just one exposure.

\begin{table} 
\caption{Inter-exposure streak-linking results for the test set with \texttt{XGBoost}. 
}
\label{table:mstable1}
\centering
\begin{tabular}{lcc}
\hline\hline
Type &XGBoost n &XGBoost \% \\
\hline
Clean hits &1025 &59.1\% \\
Semi-hits (strong) &26 &1.5\% \\
Semi-hits (weak) &26 &1.5\% \\
Clean hits + strong semi-hits &1051 &60.6\% \\
Total hits &1077 &62.1\% \\
Duplicates &281 &16.2\%\\
Misses &658 &37.9\% \\
False positives &65 &4.6\% \\
\hline
\end{tabular}
\tablefoot{
The percentages for clean hits, semi-hits, and total hits show completeness; for duplicates and false positives, they show the fraction of all detected multi-streaks that belong to that group, and for misses, they show the miss rate or false negative rate. There were 1735 ground-truth multi-streaks in the test set.\\
}

\end{table}

\begin{table}
\caption{Detailed detection completeness of the inter-exposure streak linking.}
\label{table:mstable2}
\centering
\begin{tabular}{lcc}
\hline\hline
\emph{GT multi-streaks with 4 parts} &411 &23.7\% of GTs \\
\quad of which DL found &248 &60.3\% \\
\qquad with 4 hits &207 &50.4\% \\
\qquad with 3 hits &27 &6.6\% \\
\qquad with 2 hits &14 &3.4\% \\
\emph{GT multi-streaks with 3 parts} &680 &39.2\% of GTs \\
\quad of which DL found &416 &61.2\% \\
\qquad with 3 hits &356 &52.4\% \\
\qquad with 2 hits &60 &8.8\% \\
\emph{GT multi-streaks with 2 parts} &644 &37.1\% of GTs \\
\quad of which DL found &357 &55.4\% \\
\qquad with 2 hits &357 &55.4\% \\
\hline
\end{tabular}
\tablefoot{
The numbers are calculated for clean hits. "GT" refers to ground truth and "DL" refers to deep learning, i.e., CNN+RNN+\texttt{XGBoost}.}
\end{table}

\section{Discussion}

The performance of the \texttt{StreakDet} pipeline \citep{pontinen2020} forms the baseline to which we compare the results obtained with the deep-learning pipeline. When analyzing single streaks in individual exposures, that is, before linking streaks between exposures, both completeness and purity of the deep-learning pipeline are higher than those of the \texttt{StreakDet} pipeline. The total completeness for single-exposure streaks is not explicitly stated in the \texttt{StreakDet} pipeline article, but comparing the corresponding heatmaps (Fig.~\ref{fig:Heatmap_mode2} in this paper and Fig.~3 in \citealt{pontinen2020}), we see that CNN and RNN can detect approximately 0.5 magnitudes fainter streaks than \texttt{StreakDet}, and also shorter streaks. The purity for the deep-learning pipeline at the single-exposure stage is only 6.4\%, but it actually surpasses \texttt{StreakDet}'s purity of approximately 3\%.

The deep-learning pipeline results have approximately one order of magnitude higher error for the streak angles than \texttt{StreakDet} \citep[figure~4 in][]{pontinen2020}. For example, in the magnitude range of 20--21, containing the brightest streaks analyzed, the average angle error is \ang{1.1}, and the median angle error is \ang{0.5}, compared to \ang{0.06} and \ang{0.03}, respectively, achieved by \texttt{StreakDet}. For the entire magnitude range, the average angle error is \ang{1.6}, and the median angle error is \ang{0.7}, whereas the numbers for \texttt{StreakDet} are \ang{0.2} and \ang{0.08}, respectively.

Similarly to errors on streak angles, also errors on streak lengths and streak coordinates are higher for the deep-learning pipeline than for \texttt{StreakDet}. For streaks brighter than 21 magnitudes, the deep-learning pipeline's average length error is $-2.4$\%, and the median error is $-0.5$\%, compared to $-0.6$\% and 0.04\%, respectively, for \texttt{StreakDet}. For the same set of streaks, the average coordinate error is 2.4 pixels, and the median coordinate error is 0.6 pixels, corresponding to 240 and 60 milliarcseconds, respectively. The corresponding values for \texttt{StreakDet} are 0.6 and 0.1 pixels, and 60 and 10 milliarcseconds, respectively. Across the complete magnitude range tested, the average coordinate error with deep learning is 3.8 pixels, and the median coordinate error is 0.8 pixels, while \texttt{StreakDet} achieved an average error of 1.9 pixels, and a median error of 0.2 pixels. Analyzing only streaks with length errors of less than 0.1 pixels, the average coordinate error across the magnitude range considered decreases to 0.7 pixels and the median error to 0.6 pixels, while the respective values for \texttt{StreakDet} pipeline are 0.2 pixels and 0.1 pixels.

The overall completeness of the deep-learning pipeline is higher than that of the \texttt{StreakDet} pipeline. After the CNN, RNN, and \texttt{XGBoost}, the overall clean completeness reaches 59.1\%, whereas the corresponding level reached by \texttt{StreakDet} was 55.7\%. By accepting semi-hits, which refer to detected multi-streaks containing both true and false-positive parts, the deep-learning pipeline's completeness increases to 62.1\% while \texttt{StreakDet} reached 55.9\%. Comparing the relevant completeness heatmaps \citep[Fig.~\ref{fig:HeatmapMS} in this paper and figure~5 in][]{pontinen2020}, we see that the complete deep-learning pipeline can detect asteroids that are fainter by 0.25--0.5 magnitudes compared to \texttt{StreakDet}. We also note that the deep-learning pipeline can detect shorter streaks than \texttt{StreakDet}.

Furthermore, the asteroid size distribution follows a power law, meaning that there are increasingly more faint asteroids than bright ones. Therefore, detecting streaks 0.25--0.5 magnitudes fainter would result in a greater boost in completeness than our results with a uniform simulated streak sample suggest. For example, assuming a typical power law index of 0.35 for the absolute-magnitude distribution \citep{jedicke2015}, reaching 0.5 magnitudes fainter would increase the number of detected asteroids by approximately 50\%.

\texttt{StreakDet}'s advantage in the angle and coordinate accuracies over the deep-learning pipeline is explained mainly by \texttt{StreakDet}'s dedicated PSF fitting procedures, which the software uses to estimate the streak parameters precisely. In contrast, the deep-learning pipeline does not explicitly incorporate such PSF fitting techniques. In addition to more accurate coordinate prediction, \texttt{StreakDet} has a few other advantages over deep-learning methods. Firstly, \texttt{StreakDet} can more readily be used for other surveys without retraining. In other words, it exhibits relatively high robustness, although some parameter tuning is still typically required to reach optimal performance on specific data. In contrast, the performance of deep-learning models often suffers in the case of a domain shift, which occurs when the model is tested on data following a different distribution than the training set. In other words, deep-learning models often have low robustness. We expect that the robustness of our deep-learning model is not very good and that the model will require at least some real training data in addition to the simulated data to perform well for real \textit{Euclid} images. Secondly, \texttt{StreakDet} has a more interpretable decision-making process than deep learning. Deep learning models typically work as black boxes whose inner workings are hard to interpret.

One thing to note is that the test set used in this work is more challenging in terms of linking streaks between exposures than the data used by \citet{pontinen2020}. For the \texttt{StreakDet} pipeline, all the simulated data could be used for testing. Here, we had to divide it into training, validation, and test sets. To have test-set images from all 11 simulated datasets, we chose to use certain CCDs from all the datasets for the test set. This results in fewer simulated SSOs visible in all four exposures because a large portion of the SSOs moves outside the test set CCDs between exposures. In other words, in the \texttt{StreakDet} test set, there was a much larger number of SSOs that were visible in all four exposures (41.2\%), and a relatively small number were visible in only two exposures (9.2\%). The situation has flipped for our test set, as seen in Table~\ref{table:mstable2}, so that in our test set, only 23.7\% of SSOs are visible in all four exposures, and 37.1\% are visible in only two. This change makes the linking harder because, for objects appearing four times, it is possible to detect them even if only two or three of the individual streaks were detected. However, for objects appearing in only two exposures, both streaks must be detected for them to be linked. This effect is visible in Table~\ref{table:mstable2}, where the completeness is 61.2\% for multi-streaks with three ground-truth parts and 60.3\% for four parts, whereas for multi-streaks with two parts, it is only 55.4\%. In other words, the deep-learning pipeline surpasses the completeness of the \texttt{StreakDet} pipeline even with a more demanding test set. Therefore, these estimates offer a kind of lower bound, and the deep-learning pipeline is expected to perform better on complete CCD mosaics.

Qualitative comparison to other object-detection pipelines tested on other relevant datasets shows that the performance of our pipeline appears quite strong. Also, choosing a customized streak-focused object-detection approach instead of a bounding-box-based one seems better for streak detection. \citet{varela2019} tested the YOLOv2 object-detection model \citep{redmon2017} to find satellite streaks in the data of their Wide-Field-of-View (WFoV) system. YOLOv2 reached a completeness of only 40.4\% on their human-labeled dataset. As humans labeled the dataset, the upper limit for completeness was realistically 100\% because all the streaks were visible to the human eye. Furthermore, the authors also commented on the limitations caused by the rectangular bounding boxes on the accuracy of streak coordinates. Similarly, \citet{kruk2022} tested a bounding-box-based approach using Google Cloud AutoML to detect asteroids in \textit{Hubble} images. Their AutoML model achieved 58.2\% completeness and 73.6\% purity when detecting asteroid streaks, again on a dataset where humans had reached 100\% completeness. Also, due to the bounding-box-based model, they had to develop an additional processing step to find the streak coordinates from the bounding box area. Granted, both projects were working with data for which linking streaks between exposures was not viable, meaning they had to reach much higher purity for single images than our deep-learning pipeline does, which limits the achievable completeness.

\section{Conclusions and future work}

Our deep-learning pipeline achieves similar purity and better completeness compared to the \texttt{StreakDet} pipeline tested by \citet{pontinen2020}. The deep-learning pipeline can detect shorter and fainter streaks than \texttt{StreakDet}, resulting in up to a 50\% increase in the number of detected SSOs. Therefore, it can offer more valuable scientific data on asteroids, giving insights into the nature of asteroid populations. Especially measurements of small near-Earth asteroids could prove helpful, as we currently have very little data about them. The achieved accuracy of the deep-learning pipeline for streak coordinates falls behind that of the \texttt{StreakDet} pipeline, but appears good enough to fulfill the requirements of the streak-detection task. Photometry and astrometry of the asteroids detected will be extracted after the detection step described here, with separate tools tailored for those tasks. One alternative is using the tools available in the \texttt{StreakDet} pipeline.

There is still some room for improvement in detection completeness, mainly in the last stage of the pipeline, that is, in the streak linking. A limitation of using the \texttt{XGBoost} model is that during the classification of the multi-streak candidates, the model has no information about the surrounding other streaks in the images, which in some cases could provide helpful information regarding the classification. Therefore, a more sophisticated streak-linking algorithm between the four exposures that considers all the streaks and their context in the images simultaneously, such as a graph neural network (GNN), could probably further improve both completeness and purity. Additionally, combining the NISP-P data with the VIS data could reduce false positives by adding more streaks or points along each SSO's observed arc, although due to the shorter observation time and different pixel scale, most SSOs appear as point sources in NISP-P images.

Even though our deep-learning pipeline surpasses \texttt{StreakDet}'s completeness, it may be advantageous to use both methods for \textit{Euclid}. First, they may detect partially different streaks, increasing total completeness. Second, there is typically a gap between simulated and real images, known as a domain shift; thus, a deep-learning pipeline trained only with simulated data may underperform with real data. Combining real and simulated astronomical images in the training set of a deep-learning model has shown to be a good approach, especially in terms of increasing the robustness of the model \citep{holzschuh2022}. Therefore, a sensible approach would be first to use \texttt{StreakDet} to find initial asteroid streaks from real \textit{Euclid} images, add them to our simulated training set, train the models, and then use them to find more asteroids from the images. 

Furthermore, it might be advantageous to train the deep-learning pipeline with a dataset following more realistic streak angle and length distributions, which depend on the SSO orbits and the observational geometry. The deep-learning pipeline operates in pixel coordinates, so getting the training distribution right requires considering further factors, such as the telescope orientation at different pointings and the fact that long streaks appear in more CNN cutout images than short streaks. The streak lengths in our simulated dataset cover most of the expected range of SSO streak lengths, but fast near-Earth asteroids can form longer streaks than are included in our training set, so some training examples of them should probably be added to the training set before deploying the models for the real data. More realistic angle and length distributions could be achieved by adding real asteroid streaks detected by \texttt{StreakDet} to the training set, using data augmentation to increase the proportion of more typical streak angles and lengths, generating new simulated data, or some combination thereof.

Our pipeline could be adapted relatively easily for other surveys where asteroid streaks can be linked between consecutive observations, such as the Kilo-Degree Survey \citep[KiDS;][]{saifollahi2023}. For single-exposure streak-detection purposes, our approach would have to be modified to offer a high enough purity without the inter-exposure linking stage.

\begin{acknowledgements}

We wish to thank Marcel Popescu for offering constructive comments. We acknowledge funding from the Vilho, Yrjö, and Kalle Väisälä Foundation and the Academy of Finland (projects \#316292 and \#299543). We thank the Finnish Computing Competence Infrastructure (FCCI) for supporting this project with computational and data storage resources.

\AckEC

\end{acknowledgements}

\bibliographystyle{aa} %
\bibliography{References.bib}

\end{document}